\documentclass[
superscriptaddress,
 amsmath,amssymb,
 aps, physrev,
pra,
]{revtex4-2}
\usepackage{xcolor}
\usepackage{comment}
\usepackage{physics}
\usepackage[nameinlink]{cleveref}
\usepackage{soul}
\usepackage{mathtools}
\usepackage{tikz}
\usepackage{quantikz}
\usepackage{placeins}

\usepackage{xspace}

\newcommand{\ethan}[1]{{\color{teal}Ethan:#1}}

\DeclareMathAlphabet{\mathbbold}{U}{bbold}{m}{n}

\begin{document}

\title{Quantum dynamics of spin systems using quantum-like resources}

\author{Debadrita Saha}
\affiliation{Department of Chemistry, Princeton University, Princeton, NJ 08544, USA}

\author{Ethan Dickey}
\affiliation{Department of Computer Science, Purdue University, West Lafayette, Indiana 47907, USA}

\author{William Horvat}
\affiliation{Department of Chemistry and Biochemistry, University of California, Los Angeles, CA 90095, USA}

\author{Sabre Kais}
\affiliation{Department of Electrical and Computer Engineering, Department of Chemistry, and Department of Physics, North Carolina State University, Raleigh, North Carolina 27606, USA}

\author{Gregory D. Scholes}
\affiliation{Department of Chemistry, Princeton University, Princeton, NJ 08544, USA}
\date{\today}



\begin{abstract}
Recent publications have introduced the concept of quantum-like (QL) bits, along with their associated QL-states and QL-gate operations, emerging from the dynamics of complex synchronized networks. The present work builds on this framework to present a simulation of quantum dynamics in spin systems. We focus on the transverse field Ising model, a widely studied Hamiltonian for interacting spin systems. We show how circuits, built using the QL analog of quantum gates, can be used to prepare an arbitrary initial wavepacket and simulate the time evolution of the resulting state for extracting physical observables such as magnetization. Beyond the statevector dynamics, we introduce a set of metrics that characterize how QL-circuit operations transform the underlying graph connections. These metrics quantify the microscopic coupling, phase coherence, and effective inter-block coupling generated by the circuit. 
These graph-level quantities provide constraints and design targets for potential physical implementations, indicating where interconnection resources, phase control, and effective coupling strength must be engineered in a hardware realization. Overall, this study demonstrates how classical resources can be harnessed to simulate quantum systems and we discuss the advantages and limitations. 
\end{abstract}

\maketitle

The collective behavior of individual nodes in a complex network can give rise to global states and functions that are qualitatively different from the behavior of its constituent components~\cite{hopfield_neuralnets_1982}. The brain, for example, is a network of individual neurons whose collective dynamics support computation, memory, and decision-making~\cite{bassett_network_2017}. Similar phenomena occur across widely different physical scales: local interactions between birds or fish can produce coordinated collective motion~\cite{vicsek_novel_1995}, while the spread of an epidemic can be understood as a dynamical process propagating through a network of contacts~\cite{pastor-satorras_epidemic_2001}. These examples illustrate a general principle: local interactions, when organized through a network, can generate collective states with functionality not apparent from the individual constituents. This principle has motivated the design of artificial many-body systems whose collective dynamics can be engineered to perform useful tasks, with artificial neural networks providing a prominent example.

This perspective of computation as an emergent property of collective dynamics underlies several approaches to unconventional and brain-inspired computing. In oscillator-based computing, for example, information is encoded in and processed through the collective dynamics of coupled nonlinear oscillators~\cite{csaba_coupled_2020}. Reservoir computing is a related paradigm in which a network of nonlinear dynamical elements transforms an input into a high-dimensional collective state that can subsequently be used for computation~\cite{jaeger_harnessing_2004}. Other brain-inspired approaches like memcomputing relies on non-linear dynamical systems to simultaneously process and store information for solving optimization and NP-hard problems~\cite{traversa_dcram_2014}, while neuromorphic computing implements networks of neuron- and synapse-like elements to perform low-power information processing~\cite{mead_neuromorphic_1990}. Although these approaches differ substantially in their physical realization, they share the underlying idea that computation need not be encoded in isolated logical elements, and that it can instead emerge from the collective state of an interacting system.

A natural mathematical abstraction of such systems is a graph, with the constituent elements represented by vertices and their interactions by edges. This representation separates the detailed physical implementation from the structure of the interactions and allows one to ask more general questions about how network topology controls collective behavior. In particular, can the topology of a network, together with the rules governing interactions between its nodes, determine the existence and properties of robust collective states? And, if the network can be engineered to generate such states, can information be encoded in the collective state itself? Such an encoding would distribute information across the global structure of the network while retaining the possibility of manipulating it through local operations. This provides a natural route from network dynamics to a computational architecture in which collective states serve as the computational degrees of freedom.

Quantum-like (QL) computing follows this idea by using classical networks whose synchronized states are designed to reproduce selected features of quantum mechanical state spaces~\cite{scholes_quantumlike_2024,scholes_qlcartpdt_2025}. In this framework, the network topology and coupling are engineered so that the emergent collective states mimic those of quantum many-body systems. The foundations of QL-based computing were introduced in Ref.\cite{scholes_quantumlike_2024}, where random graphs with expander-like connectivity were coupled to generate collective states corresponding to many-body quantum states. Furthermore, QL analogs of standard quantum gate operations were introduced in Ref.~\cite{amati_quantum_2025} to enable encoding and processing of information using these QL resources. The resulting framework therefore provides a classical network representation of a quantum state space, while retaining the possibility of implementing operations through the dynamics of the underlying network. In Section~\ref{sec:qlcircuits}, we provide a primer on QL-bit construction and describe how circuit model quantum computing operations can be represented in the language of QL-bits.

The connection to quantum many-body systems is particularly relevant because simulating their dynamics is one of the central computational challenges across condensed-matter physics, atomic and molecular physics, chemistry, and materials science. Classical simulations are fundamentally limited by the exponential growth of the many-body Hilbert space, motivating methods that exploit structure in physically relevant states or a physical computational architecture whose state space naturally scales exponentially. Exact diagonalization provides a benchmark for small systems, while tensor-network methods, including matrix-product states and projected entangled-pair states, exploit the limited entanglement of broad classes of physically relevant states to obtain efficient approximate representations~\cite{verstraete_matrix_2008,orus_mpspeps_2014}. 

Quantum computing provides an alternative approach to many-body simulation by encoding the many-body state in an exponentially large quantum Hilbert space. Quantum simulation methods can be broadly classified as analog and digital approaches, with both having been used to study spin models, many-body localization, lattice gauge theories, and nonequilibrium dynamics across platforms including trapped ions, ultracold atoms, and superconducting circuits~\cite{fauseweh_quantum_2024}. Despite these advances, current quantum hardware remains limited by noise, decoherence, finite circuit depth, and the overhead associated with implementing and controlling many-body dynamics~\cite{fauseweh_quantum_2024}. These limitations motivate the exploration of alternative physical architectures for representing and manipulating many-body quantum states. 

These complementary approaches motivate an intermediate question: can a classical physical architecture encode and manipulate a computational state space whose dimensionality scales exponentially? QL computing addresses this question by using collective states of interacting classical networks as an alternative representation of quantum-like many-body states. In this work, we investigate this possibility by using the graph formulation of complex classical networks to simulate the dynamics of a quantum many-body system. Specifically, in Section~\ref{sec:model}, we construct a network whose collective QL states reproduce the state space of a lattice of spins and use its dynamics to simulate the corresponding quantum evolution. The resulting quantum statevector simulations are presented in Section~\ref{sec:results}.

A separate question is whether networks with the required connectivity can be realized as physical systems. This issue is particularly important because the network structures used to generate robust collective states can require highly nonlocal connectivity. The difficulty of embedding such connectivity in physical space was recognized in early studies of expander graphs motivated by models of neural networks~\cite{kolmogorov_realization_1993}. Therefore, rather than viewing the network merely as an implementation of a particular algorithm, we examine how the network structure encodes the evolving quantum state. In Section~\ref{sec:metrics}, we introduce metrics for characterizing how the network structure changes during the dynamics simulation. The resulting network-dynamics simulations are presented in Section~\ref{sec:adj_dyn}.

This distinction between the mathematical network and its physical realization motivates the discussion in Section~\ref{sec:outlook}. The synchronization and collective-state properties considered here are properties of idealized graph models; whether physically realizable systems can reproduce the required connectivity, synchronization, and robustness at scale remains an open question. We therefore do not claim to resolve the physical implementation problem in this work. Instead, we identify physical realization as an important direction for future development of QL computing and discuss the constraints that must be addressed to translate the graph-based construction into a realizable computing architecture.

\begin{figure}
    \centering
    \includegraphics[width=\linewidth]{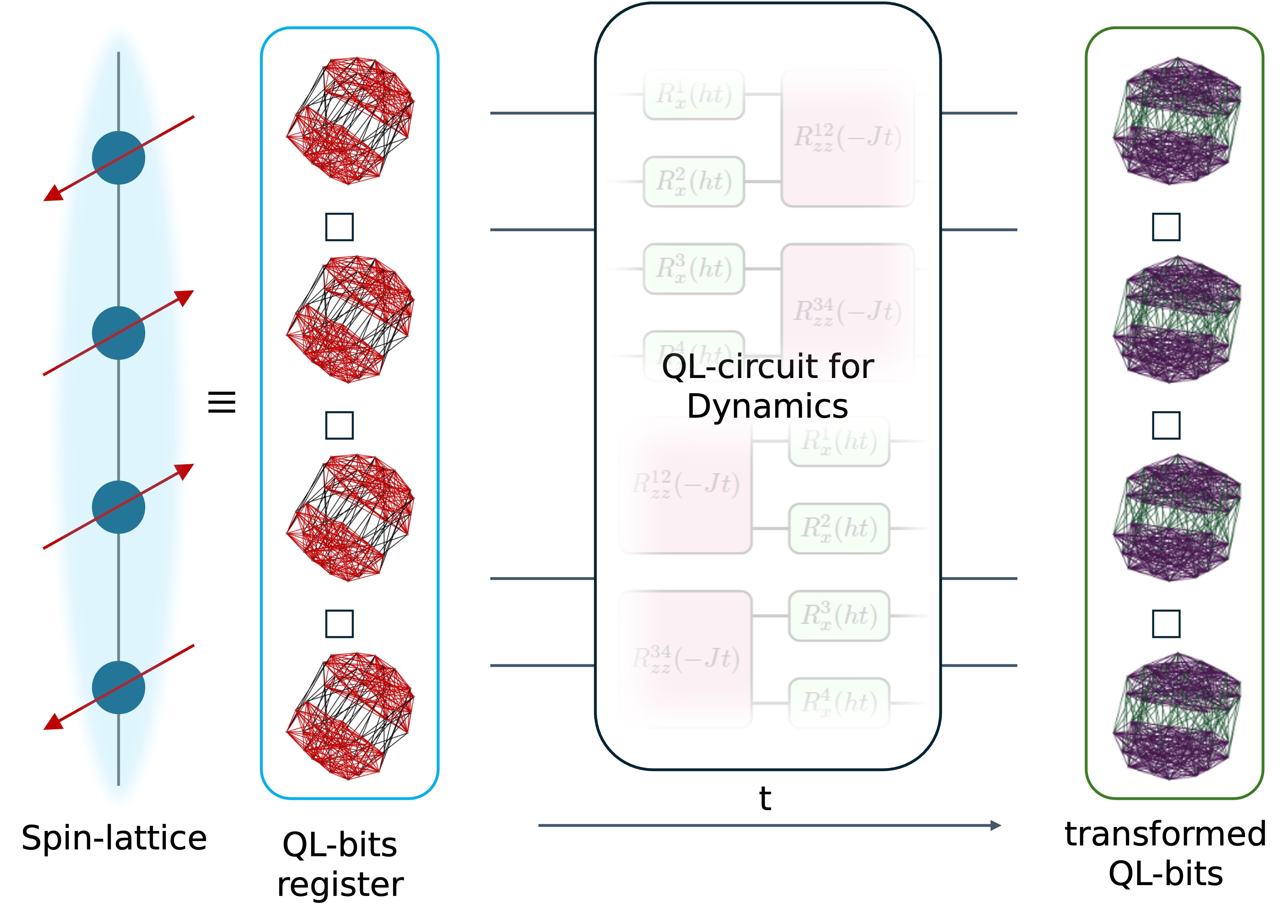}
    \caption{Schematic overview of the QL-bit representation of spin dynamics. A one-dimensional spin lattice (\textbf{left}) is represented as a network of QL-bits. The initial state of the spin system is encoded in the emergent state of the Cartesian product of QL-bits. Time evolution of the spin system is implemented as a QL-circuit acting on these QL-bits, transforming their graph structure over time \textit{t} and producing the evolved QL-bit representation (\textbf{right}). 
    }
    \label{fig:overview}
\end{figure}

\section{Theoretical Background}\label{sec:qlcircuits}
The foundation for the QL-bit formalism were laid in Ref~\cite{scholes_quantumlike_2024}. Ref~\cite{dickey_programmable_2026} subsequently placed this construction on a programmable footing by 
showing that detuned subgraph regularities or asymmetric intergraph couplings can realize arbitrary real-amplitude single QL-bit states (with arbitrary complex-amplitude states subsequently realized using Hermitian weighted couplings in \cite{dickey2026complex}). The construction extends to QL-dits using $d$ coupled expander subgraphs~\cite{saha_multi-dimensional_2025}. Ref.~\cite{scholes_qlcartpdt_2025} introduced a general framework for constructing multi-QL-bit resources, using the Cartesian product ($\Box$) of graphs. 
Furthermore,~\cite{amati_quantum_2025} showed how arbitrary gates can be implemented by manipulating emergent states in these systems and~\cite{xie_dynamicalreadout_2026} demonstrates that a structured classical network can encode effective two-qubit states and dynamically reconstruct their measurement statistics, including CHSH correlations like entanglement that exceed the classical bound, directly from the network’s spectral response.


For the purpose of quantum information processing on the quantum-like resources, we need QL-bits to encode quantum states, elementary unitary operations to build arbitrary unitary operations on the states, and a measurement framework to compute expectation values of observables and extract desired properties. In this section, we discuss prior work and advances on all these fronts within the QL-framework.

\subsection{QL States}\label{subsec:states}
We construct a single QL-bit as a graph, $\mathcal{G}^{(q)}$, composed of two distinct subgraphs, $\mathcal{G}^{(q)}_{A}$ and $\mathcal{G}^{(q)}_{B}$ (as shown in Ref.~\cite{scholes_quantumlike_2024}). Each subgraph, $\mathcal{G}^{(q)}_{A}$ and $\mathcal{G}^{(q)}_{B}$, is constructed as an undirected $k$-regular graph. Each node in $\mathcal{G}^{(q)}_{A}$ is connected to $\ell$ nodes in $\mathcal{G}^{(q)}_{B}$. (See Figure~\ref{fig:overview}, labeled `QL-bits register': the subgraphs have edges shown in red, while inter-graph couplings are shown in black.) The corresponding adjacency matrix $\mathcal{R}^{(q)}$, for the QL-bit graph $\mathcal{G}^{(q)}$ has the following block-structured form:
\begin{align}\label{eqn:ql-bit}
    \mathcal{R}^{(q)} = 
    \begin{bmatrix}
        A_{A} & C \\
        C^{T} & A_{B}
    \end{bmatrix}
\end{align}
where the diagonal blocks $A_{i}$, $i \in \{A,B\}$ represent the adjacency matrices of the subgraphs, and $C$ represents the coupling between the subgraphs $\mathcal{G}^{(q)}_{i}$. The two largest eigenvalues of $\mathcal{R}^{(q)}$, $\lambda_\pm = k\pm \ell$, are isolated from its incoherent state spectrum (or the bulk spectrum) and their corresponding eigenstates form a basis for the two-dimensional QL-bit state space.

We employ the Cartesian product of graphs to form composite QL-resources comprised of multiple QL-bits~\cite{scholes_qlcartpdt_2025}. The Cartesian product of $N_{QL}$ such QL-bit graphs is given by $\mathcal{G} = \mathop{\Box}\limits_{q=1}^{N_{QL}} \mathcal{G}^{(q)}$, with the corresponding adjacency matrix $\mathcal{R} = \sum_{q=1}^{N_{QL}} \left[ \bigotimes_{p=1}^{q-1} \mathbbold{1}_{2n} \right] \otimes \mathcal{R}^{(q)} \otimes \left[ \bigotimes_{p' = q+1}^{N_{QL}} \mathbbold{1}_{2n} \right]$  is expressed in terms of the individual adjacency matrices $\mathcal{R}^{(q)}$ of the QL-bits,
with $n$ number of vertices in each subgraph of $\mathcal{G}^{(q)}$. The Cartesian product of individual QL-bits has been shown to map to the tensor product of Hilbert spaces corresponding to quantum bits~\cite{scholes_qlcartpdt_2025}. From this construction, the eigenvalues of $\mathcal{R}$ can be expressed as the sums of the emergent eigenvalues $\lambda_{i}^{(q)}$ of the constituent QL-bits. The corresponding emergent eigenstates of the product graph are tensor products ($\bigotimes_{i} \phi_{i}$) of the emergent states ($\phi_{i}$) of the constituent QL-bits.

For multi-QL-bit resources, the Cartesian product construction incurs an exponential cost of $(2n)^{N_{QL}}$, with respect to the number of vertices in each subgraph. To reduce the resource scaling while preserving the emergent eigenstates which we use to represent our states, we rely on the formalisms and constructions of Ref.~\cite{horvat_minimal_2026}, which introduced the minimal representation of QL-bits.

The minimal representation is generated by performing an equitable partition, $\pi(\mathcal{G}^{(q)})$, on the vertices of the graph $\mathcal{G}^{(q)}$. The resulting coarse grained graph is a quotient graph with adjacency matrix denoted by $\mathcal{R}_{\pi}$. The relationship between the full adjacency matrix and the quotient adjacency matrix is given by:
\begin{equation}\label{eqn:r2rpi}
    \mathcal{R}_{\pi} = (S^T S)^{-1} S^T \mathcal{R} S
\end{equation}
where $S$ is the characteristic matrix of the equitable partition.

\subsection{QL Gates}\label{subsec:gates}
The QL-states encode information that can be manipulated using transformations like quantum gates in conventional quantum computing circuits. For that, Ref.~\cite{amati_quantum_2025} introduces a map $U_{cb} = [V_{H}\otimes\mathbbold{1}_{n}]^{\otimes N_{QL}}$ that transforms the emergent eigenvectors into the standard factorized computational basis, where $V_H$ is the standard Hadamard gate in quantum computation. The QL analog, $U_g$, of an arbitrary quantum gate $V_g$ on a single QL-bit is defined by the matrix transformation,
\begin{equation}\label{eqn:Ugdef}
    U_g = U_{cb}^{-1} (V_{g}\otimes \mathbbold{1}_{n})U_{cb} = \widetilde{V}_{g}\otimes \mathbbold{1}_{n}
\end{equation}
where we define $\widetilde{V}_{g} = V_{H}^{-1}V_{g}V_{H}$. This can be extended to define gate operations for arbitrary number of QL-bits. For example, the QL analog of CNOT gate that operates on two QL-bits can be defined as $U_{CN} = U_{cb}^{-1} (\ket{0}\bra{0}\otimes \hat{I} + \ket{1}\bra{1}\otimes \hat{X})U_{cb}$~\cite{amati_quantum_2025}, where $\hat{I}$ and $\hat{X}$ are the identity and Pauli $X$ operators, respectively.

Using the general recipe for defining QL-gates, we use \Cref{eqn:Ugdef}  to define gates with continuous rotation angles about the \textit{x}, \textit{y}, \textit{z} directions, $R_x$, $R_y$, and $R_z$ gates. The explicit forms of the QL-analogs of these gates are provided in the SI Section 1.

The unitary operation $U_{g}$ corresponding to a specific gate $g$ transforms the resource $\mathcal{R}$ of a single QL-bit via $\mathcal{R}_{g} = U_{g} \mathcal{R} U_{g}^{\dagger}$. All gate operations on the full graph adjacency matrix ($\mathcal{R}$) can be performed in an equivalent manner on the quotient graph adjacency matrix. The characteristic matrix, $S$, that affects the transformation $\mathcal{R}\rightarrow \mathcal{R}_\pi$ in \Cref{eqn:r2rpi} aids the gate transformations to the minimal representation. Following the formal derivation in SI Section 2, 
one arrives at the following gate operation equivalence,
\begin{align}
    (\mathcal{R}_g)_{\pi} &= \widetilde{V}_{g} \mathcal{R}_{\pi} \widetilde{V}_{g}^{\dagger}
\end{align}
where $\widetilde{V}$ is the gate operation in the minimal representation. We use $g$ to represent either one single-QL-bit gate or a sequence of single- and two-QL-bit gate operations, as we will see later on in the paper.

\subsection{QL-circuits}\label{subsec:ql-circuits}

A QL-circuit is a sequence of QL-gate operations applied to one or more QL-bits, analogous to how a quantum circuit sequences gates on qubits. The initial state, $\ket{\psi(0)}$, of the system can be encoded in the emergent state of the QL-resource, either a single QL-bit or the Cartesian product of $N_{QL}$ QL-bits.
Following the conventional circuit-based method, one can compute the unitary that transforms $\ket{000\ldots 00}$ into the desired initial state and construct the corresponding QL-circuit.

A general QL-circuit consists of layers of gate operations, $u_i$, applied in sequence, with each layer built from tensor products of single- and two-QL-bit gates. We use the gate set $\{R_x, R_z, \mathrm{CNOT}\}$ which is universal for quantum computation, since arbitrary single-qubit unitaries can be synthesized via Euler-angle decompositions and CNOT provides entanglement. While Clifford-only circuits can be classically simulated, arbitrary-angle rotations in $R_x$ and $R_z$ are generally non-Clifford, which ensures (at least for qubits) that the set achieves universality and can be used to decompose any circuit operation~\cite{barenco_elementary_1995}. Several efficient unitary decomposition methods exist and, in principle, the QL-formalism does not constrain us to a particular one. 

We can decompose the unitary time evolution operator $\mathcal{U}(\delta t)$ corresponding to the Hamiltonian $\mathcal{H}$ of a system into $M$ layers of unitaries, $\mathcal{U}(\delta t) = \prod_{i=1}^{M} u_{i}$, giving the QL-circuit analog of a quantum time evolution, $\ket{\psi(t)} \to \mathcal{U}(\delta t) \ket{\psi(0)}$. Each layer $u_{i}$ further consists of tensor products of only single- and two-QL-bit gate operations.

When simulating dynamics using a QL-circuit, $U_{t}$ denotes the accumulated QL unitary at a time step $t$, and the corresponding transformed graph adjacency matrix is $\mathcal{R}_{t} = U_{t} \mathcal{R}_{0} U_{t}^{\dagger}$. Layers composed of independent single QL-bit gates preserve the Cartesian product form derived above. The two QL-bit interaction layers generally do not, and can instead produce a more general complex Hermitian weighted graph operator. We define $U_{t}$ in Section~\ref{subsec:dynamics}. Since this is a unitary transformation, the spectrum of $\mathcal{R}_{t}$ are preserved.

\section{Ising Model and Dynamics}\label{sec:model}
In this section, we apply the QL-circuit formalism developed above to simulate the quench dynamics of the transverse-field Ising model. We first define the model and the quench protocol (Section~\ref{subsec:Ising}), then describe the Trotterized decomposition used to construct the time-evolution circuit (Section~\ref{subsec:dynamics}), and finally define the observables used to characterize the post-quench dynamics (Section~\ref{subsec:computing_observables}).

\subsection{Transverse Field Ising Model}\label{subsec:Ising}
We consider here a one-dimensional finite spin chain of \textit{L} spin-$\frac{1}{2}$ operators ($\sigma_{x}^{i},\sigma_{y}^{i},\sigma_{z}^{i}$) that is popularly used across condensed matter physics and atomic, molecular and optical physics. (Refer to Figure~\ref{fig:overview}, left panel for schematic.) Moreover, since qubit states can be modeled as spin states, Ising Hamiltonians, in general, have found wide application in various implementations of quantum computing algorithms. In particular, we study the transverse field Ising (TFI) model, with open boundary conditions,  with  Hamiltonian,
\begin{align}
    H_{TFI} = -J\sum_{\langle i,j\rangle}\sigma_{i}^{z}\sigma_{j}^{z}+h\sum_{i}\sigma_{i}^{x},   
\end{align}
where $\langle, \rangle$ denotes nearest neighbor interactions, $J$ is the spin-spin coupling interaction and $h$ is the transverse magnetic field strength. The TFI model is a paradigmatic testbed for quantum many body physics. It is an exactly solvable model using the Jordan-Wigner transformation~\cite{pfeuty_tfi1D_1970}, exhibits a continuous quantum phase transition at $h_{c}=J$ from the ferromagnetic ordered to the paramagnetic phase~\cite{sachdev_quantum_2011}, and is the original system in which dynamical quantum phase transitions were observed~\cite{heyl_dqpt_2013}. The model has been realized experimentally on trapped-ion~\cite{kim_quantum_2010}, Rydberg-atom~\cite{bernien_probing_2017}, and superconducting circuit~\cite{dborin_simulating_2022} platforms. Its applications on quantum computing hardware range from electronic structure problems~\cite{xia_electronic_2018} to simulating dynamical quantum phase transitions~\cite{zhang_dqpt53_2017} and solving optimization problems using adiabatic quantum computation~\cite{wiebe_qdeeplearn_2016}. This range of applicability and experimental accessibility makes it a natural benchmark for validating new quantum simulation methods, including the QL-circuit framework developed here. 

We consider several cases of quench in this model, ranging from a slow quench at $h=0.2$ to a fast quench at $h = 2.0$, in intervals of $0.2$. In all cases, we assume the spin chain to initially be uniformly ordered with all spins in the $\ket{\uparrow}$ state, and we set $J = 1$.

\subsection{Trotterized Time Evolution}\label{subsec:dynamics}
There are several existing methods for simulating the time evolution operator corresponding to the Hamiltonian of a physical system. Exact diagonalization remains the gold standard for small systems but is limited to a few tens of spins even with sparse-matrix techniques \cite{lin_exact_1993}. Tensor-network methods such as the density-matrix renormalization group (DMRG) \cite{white_dmrg_1992} and its time-dependent extensions, including time-evolving block decimation (TEBD) \cite{vidal_efficient_2004} and matrix-product-state-based time-dependent variational principle (TDVP) approaches \cite{haegeman_tdvp_2011}, extend accessible system sizes considerably for low-entanglement dynamics, though they become inefficient once entanglement entropy grows extensively, as is typical after a quench \cite{schollwoeck_dmrg_2011}. On digital quantum hardware, product-formula (Trotter-Suzuki) decompositions have become the standard route to implementing many-body time evolution, following the original operator-splitting formulas of Trotter \cite{trotter_product_1959} and Suzuki \cite{suzuki_generalized_1976} and their adaptation to universal quantum simulation by Lloyd \cite{lloyd_universalqsim_1996}. 

\begin{figure}[htb!]
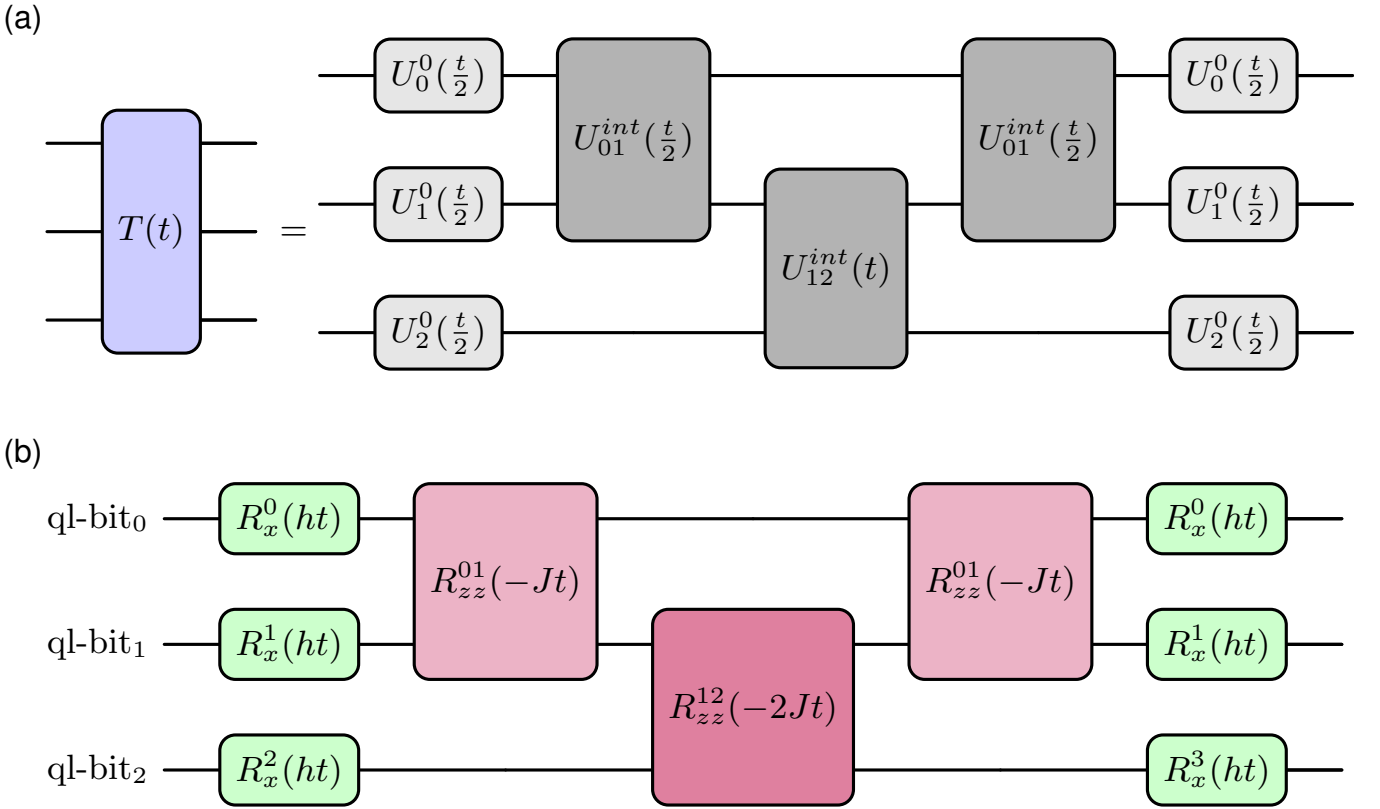

    \centering
     \makebox[\linewidth][l]{%
         \hspace{-0.1cm}%
         {\fontfamily{phv}\selectfont\fontsize{12}{16}\selectfont (a)}
     }
     \input{figures/trotterstep_n3}\\
    \makebox[\linewidth][l]{%
        \hspace{-0.1cm}%
        {\fontfamily{phv}\selectfont\fontsize{12}{16}\selectfont (b)}
    }
    \input{figures/trotterstep_n3_gates}
    \caption{(a) The second-order Trotter-Suzuki decomposition for a single step ($N=1$), $T(t)$, for a general interacting Hamiltonian. (b) Gate based implementation of the circuit for the transverse field Ising model for a one-dimensional lattice with three spin sites.}
    \label{fig:trotterstep1}
\end{figure}

In the Trotter approach, the assumption is that the Hamiltonian can be decomposed into polynomially many local operators that act only on a non-extensive number of qubits. The error in the resulting decomposition comes from the non-commutativity of these local operators. Here, we follow the ($N$-step) second order Trotter-Suzuki decomposition summarized in Ref.~\cite{buessen_distributed_2023} as uniform Trotterization. Figure~\ref{fig:trotterstep1} depicts a single step of the decomposition. The approximation error scales as $\approx N\delta t^{3}$. The TFI model Hamiltonian discussed above can be written as a general Hamiltonian given by $H = \sum_{i}H_{(i)}^{0} + \sum_{\langle i,j\rangle}H_{i,j}^{int}$, where $H_{(i)}^{0}$ denotes a local term on site $i$ and $H_{i,j}^{int}$ denotes an interaction term between sites $i$ and $j$. The terms of the Hamiltonian generally do not commute. The time evolution operator $U(t) = e^{-iHt}$ can be approximated as a sequence of one- and two-qubit operations, $U_{i}^{0}(t) = e^{-iH_{(i)}^{0}t}$ and $U_{i}^{i,j}(t) = e^{-iH_{i,j}^{int}t}$, as
\begin{equation} \label{eqn:trotter_dfn}
    T^{N}(t) = (T^{0}(\frac{\delta t}{2})T^{\text{even}}(\frac{\delta t}{2})T^{\text{odd}}(\delta t)T^{\text{even}}(\frac{\delta t}{2})T^{0}(\frac{\delta t}{2}))^{N}
\end{equation}
where ~$\delta t = \frac{t}{N}$, ~$T^{0}(t) = \prod_{i}U_{i}^{0}(t)$, and ~$T^{\text{even}}(t)=\prod_{\langle i,j\rangle}U_{i,j}^{int}(t)$, with the product running over all even pairs of nearest-neighbor sites and the nearest-neighbor pairs on the chain labeled even and odd alternatingly. $T^{\text{odd}}(\delta t)$ is defined analogously. The corresponding circuit is shown for the case of three spins in the lower panel of \Cref{fig:trotterstep1}.

\subsection{Computing Observables}\label{subsec:computing_observables}
Having constructed the time evolved state $\ket{\psi(t)}$ using the Trotter circuit discussed above, we characterize the dynamics of the system through the magnetization along the $z$ direction, the direction conjugate to the $\mathbb{Z}_{2}$ symmetry broken in the ferromagnetic phase of the TFI model. The site-resolved magnetization in the direction perpendicular to the transverse magnetic field at time $t$ is,
\begin{align}
    M_{z}^{i}(t) = \bra{\psi(t)}\sigma_{z}^{i}\ket{\psi(t)},
\end{align}
and the spatially averaged magnetization at each time step is:
\begin{align}
    M_{z}(t) = \frac{1}{N}\sum_{i} M_{z}^{i}(t)
\end{align}
Following a quench, $M_{z}(t)$ generally exhibits persistent oscillations rather than relaxing to a fixed value, so we characterize the long-time behavior through time averaged magnetization over a finite window $T$:
\begin{align}
    {\bar{M_{z}}} = \frac{1}{T} \int_{0}^{T} M_{z}(t) dt
\end{align}
This provides a qualitative form of an order parameter for detecting the underlying quantum phase transition in the post-quench dynamics.

\section{Graph Dynamic Variables}\label{sec:metrics}
The QL-circuit framework admits two complementary descriptions of the dynamics of the system, the evolution of the statevector $\ket{\psi(t)}$, and equivalently, the corresponding transformation of the underlying graph adjacency matrix, $\mathcal{R}_{t} = U_{t} \mathcal{R}_{0} U_{t}^{\dagger}$. 
Section~\ref{subsec:computing_observables} characterized the former through $\bar{M}_{z}$. Here, we turn to the latter, introducing a set of graph-level metrics that quantify how the Trotter circuit redistributes coupling across the Cartesian product structure over time. These diagnostics help guide the practical implementation of our QL-circuit formalism. To aid our analysis, we introduce the following block structure on the Cartesian product of graph over $N_{QL}$ QL-bits.

The vertices of the full Cartesian product can be partitioned into $d=2^{N_{QL}}$ blocks indexed by strings in $\{A,B\}^{N_{QL}}$ corresponding to the labels for subgraphs $\mathcal{G}_{A}$ and $\mathcal{G}_{B}$ of the QL-bits that the vertices belong to. We relabel and denote these blocks by $C_1,\ldots,C_d$, where the number of nodes in the $r^{\text{th}}$ block is $n_r=|C_r|$. Ordering the vertices according to this partition gives the block decomposition
\begin{equation}\label{eqn:cartesian-block-decomposition}
    \mathcal{R}_t = \left[ \mathcal{R}_{rs}(t) \right]_{r,s=1}^{d}, \qquad \mathcal{R}_{rs}(t)\coloneqq\mathcal{R}_t[C_r,C_s],
\end{equation}
where $\mathcal{R}_{rs}(t)$ is the submatrix with rows indexed by $C_r$ and columns indexed by $C_s$. With this ordering, the diagonal submatrices describe couplings within the individual blocks, while the off-diagonal submatrices describe inter-block couplings. Since $\mathcal{R}_t$ is Hermitian, $\mathcal{R}_{sr}(t)=\mathcal{R}_{rs}(t)^\dagger$.

Each Cartesian block has a normalized uniform mode $\ket{\overline{C_r}} = \frac{1}{\sqrt{n_r}}\sum_{i\in C_r}\ket{i}$. $P = \left[\; \ket{\overline{C_1}} \cdots \ket{\overline{C_d}} \;\right] = S(S^T S)^{-1/2}$,
where $S$ is the characteristic matrix of the partition introduced in \Cref{eqn:r2rpi}. Since the blocks $C_{r}$ partition the vertex set, the $\ket{\overline{C_r}}$ are mutually orthonormal, $P^\dagger P=I_d$, and $Q_t=P^\dagger\mathcal{R}_tP$ retains the coupling between the normalized uniform modes of the Cartesian blocks. For the equal-sized Cartesian blocks considered here, $Q_t$ coincides with the quotient construction in \Cref{eqn:r2rpi} applied to $\mathcal{R}_t$. More generally, it is the corresponding symmetrically normalized compression. We call this reduction \textit{synchronized compression} because each Cartesian block is represented by a single uniform mode, with all vertices within that block sharing the same amplitude and phase.

The three variables defined below distinguish the weight of coupling within and between blocks, the extent to which its complex entries cancel, and the net coupling retained in the synchronized compression.

\subsection{Microscopic Block Strength}\label{subsec:microscopic}
To quantify the total edge weight in $\mathcal{R}_{rs}(t)$ connecting the nodes of block $C_{r}$ to the nodes of block $C_{s}$, whether within a single block ($r=s$), or between two distinct blocks ($r\neq s$), independent of any cancellations among independent entries, 
we define the microscopic block strength
\begin{equation} \label{eqn:microscopic-block-strength}
    W_{rs}(t) = \left(\sum_{i\in C_r}\sum_{j\in C_s}|(\mathcal{R}_t)_{ij}|^2\right)^{1/2} .
\end{equation}
Here $i$, $j$, denote the nodes of the Cartesian product graph that belong to the blocks $C_{r}$ and $C_{s}$, and $(\mathcal{R}_t)_{ij}$ is the weight of the edge between them. Equivalently, $W_{rs}$ is the Frobenius norm of the $r-s$ block of $\mathcal{R}_t$. 
$W_{rs}$ is zero exactly when no edges connect the two blocks.

Note that, since unitary similarity preserves the Frobenius norm of the full matrix, $ \|\mathcal{R}_t\|_F^2 = \sum_{r,s}W_{rs}(t)^2$ is conserved over time. Individual $W_{rs}$ values, therefore, trace how this fixed total is redistributed across blocks as the dynamics evolves. A larger Frobenius norm can reflect more edges, larger edge weights, or both.

Looking at $N_{QL} = 2$, for example, the blocks in the Cartesian product graph are denoted by $\{AA, AB, BA, BB\}$. 
$W_{rr}$ measures the total microscopic weight represented by the internal connectivity or edge density, within the blocks $\{AA, AB, BA, BB\}$. $W_{rs}$ provides the strength of interaction between blocks $r-s$ (for example $AA-AB$).

\subsection{Weighted Phase coherence}\label{subsec:phase_coherence}
As mentioned earlier, for a nonempty block, the edge weights $(\mathcal{R}_t)_{ij}$ connecting $C_{r}$ to $C_{s}$ are, in general, complex, and can add constructively or destructively depending on their relative phase. We quantify this alignment by defining its weighted phase coherence as
\begin{equation} \label{eqn:weighted-phase-coherence}
    \chi_{rs}(t)=
    \frac{\left|\sum_{i\in C_r}\sum_{j\in C_s}(\mathcal{R}_t)_{ij}\right|}{\sum_{i\in C_r}\sum_{j\in C_s}|(\mathcal{R}_t)_{ij}|},
    \qquad 0 \leq \chi_{rs}(t) \leq 1.
\end{equation}
The numerator is the magnitude of the coherent sum of edge weights between the two blocks, while the denominator is the sum of their individual magnitudes, the largest value the coherent sum can take if all phases aligned. Note that this denominator differs from $W_{rs}$ in that it sums magnitudes directly rather than their squares, so $\chi_{rs}$ and $W_{rs}$ probe different aspects of the same block. 

$\chi_{rs}$ is, therefore, the fraction of the largest possible weighted resultant, with all entry magnitudes held fixed, that survives when the entries are added with their actual phases. Thus, $\chi_{rs} = 1$ means complete alignment (all edge weights have the same phase), $0<\chi_{rs}<1$ indicates partial cancellation (phases partially oppose), and $\chi_{rs}=0$ indicates complete cancellation (edge weights sum exactly to zero). $\chi_{rs}$ is undefined for an empty block.

\subsection{Effective Synchronized Coupling Magnitude}\label{subsec:effsync}
With the normalized Cartesian-block modes collected in $P$ as above, the synchronized compression is $Q_t=P^\dagger\mathcal{R}_tP$, whose entries are
\begin{align} \label{eqn:synchronized-compression}
    Q_{rs}(t)&=\bra{\overline{C_r}}\mathcal{R}_t\ket{\overline{C_s}}=\frac{1}{\sqrt{n_rn_s}}\sum_{i\in C_r}\sum_{j\in C_s}(\mathcal{R}_t)_{ij}. 
\end{align}
For $r \neq s$, $|Q_{rs}|$ is the effective synchronized coupling magnitude between the normalized uniform modes of $C_s$ and $C_r$. The factor $1/\sqrt{n_rn_s}$ comes from normalizing those modes; it does not make $Q_{rs}$ a probability or restrict it to $[0,1]$. The relation
\begin{equation} \label{eqn:coherence-coupling-relation}
    |Q_{rs}(t)|=\chi_{rs}(t)\frac{\sum_{i\in C_r}\sum_{j\in C_s}|(\mathcal{R}_t)_{ij}|}{\sqrt{n_rn_s}}
\end{equation}
shows why $W_{rs}$, $\chi_{rs}$, and $|Q_{rs}|$ answer different questions.
In particular, $W_{rs}$ is the $\ell^2$ norm of the block’s entries, while the right side above uses their $\ell^1$ norm before accounting for phase cancellation and normalization.
A useful interpretation of \Cref{eqn:coherence-coupling-relation} is that $Q_{rs}$ can change because the total absolute microscopic weight changes, the phase coherence changes, or both.


\Cref{tab:block-diagnostic-cases} shows possible combinations of these three parameters and associated interpretations. In particular, large $W_{rs}$ together with small $\chi_{rs}$ and small $|Q_{rs}|$ identifies a large microscopic structure that is hidden by phase cancellation in the quotient description (which we will see happen in our case, in Section~\ref{sec:adj_dyn}).


In a coupled-oscillator realization of QL-bits, the interconnection network can dominate hardware complexity, but a physical connection alone does not guarantee the intended effective dynamical coupling \cite{csaba_coupled_2020}. These three metrics connect the block description to three complementary implementation concerns.

If the entries of $\mathcal{R}_t$ are realized as programmable complex couplings, 
$W_{rs}$ quantifies the aggregate microscopic coupling weight that must be supported between Cartesian blocks $r$ and $s$. It therefore identifies where inter-block resources are required and the overall coupling scale that each block pair must support, although it does not determine whether that weight should be realized through more physical channels, stronger individual channels, or mediated interactions. 

The phase coherence $\chi_{rs}$ then indicates how carefully the phases of those microscopic couplings must be coordinated. In particular, a large $W_{rs}$ together with a small $\chi_{rs}$ means that the target operator relies on cancellation among substantial microscopic couplings, so interactions that support independent phase control are required to reproduce it. 

Finally, $|Q_{rs}|$ allows us to move from the individual graph elements to the coarse-grained synchronized dynamics. It gives the magnitude of the effective coupling that an experimental realization should reproduce between the normalized synchronized modes of blocks $r$ and $s$, while the phase of $Q_{rs}$ would be required separately to specify the full complex interaction. 
Multiple microscopic coupling patterns can produce the same $Q_{rs}$, so the quotient-level target does not uniquely determine the underlying hardware realization. Together, the three diagnostics separate the interconnect burden, the phase-control requirement, and the effective encoded coupling of a QL implementation.

\begin{table*}[t]
    \centering
    \small
    \caption{Interpretation of the block diagnostics.}
    \label{tab:block-diagnostic-cases}
    \begin{tabular}{p{0.09\textwidth} p{0.14\textwidth} 
                    p{0.20\textwidth} p{0.45\textwidth}}
        \hline
        $W_{rs}$ & $\chi_{rs}$ & Effect on $|Q_{rs}|$ & Interpretation \\
        \hline
        $0$ & Undefined & $0$ & No microscopic block is present. \\
        $>0$ & Near $1$ & Nearly maximal for the given magnitudes
        & The microscopic entries are strongly phase aligned. \\
        $>0$ & Between $0$ and $1$ & Reduced
        & The block undergoes partial phase cancellation. \\
        $\gg 0$ & Near $0$ & Near $0$
        & Strong microscopic structure is hidden by coherent cancellation in the coupling between the blocks' uniform modes. \\
        \hline
    \end{tabular}
\end{table*}

\section{State Vector Results}\label{sec:results}
We first study the unitary dynamics of the system using the QL-circuit method discussed in Section~\ref{sec:model}. The QL-bits are constructed from graphs with number of nodes $n = 32$, regularity of $k = 20$, and inter-graph connectivity $l = 5$. For the case of the transverse Ising Hamiltonian, each spin is mapped to a QL-bit, with the $\{\ket{0},\ket{1}\}$ states of the QL-bit corresponding to the $\{\ket{\uparrow},\ket{\downarrow}\}$ spin states. The emergent state of the Cartesian product graph, $\ket{000\ldots 00}$, thus corresponds to the spin state $\ket{\uparrow \uparrow \uparrow \ldots \uparrow \uparrow}$. In the quench protocol, we first prepare the initial state as the ground state, $\ket{\psi(0)} = \ket{00\ldots 0} = \ket{\uparrow\uparrow\ldots \uparrow}$, of the transverse Ising Hamiltonian, $H_{0} \equiv H(J, h_{0})$, and then suddenly turn on the transverse field which corresponds to $H_{1} \equiv H(J, h_{1})$ at $t=0$.  The subsequent time evolution is then governed by $\ket{\psi(t)} = e^{-iH_{1}t}\ket{\psi_{0}}$. The time evolution operator is computed for a time step $\delta t$ using the Trotterization method discussed in Section~\ref{subsec:dynamics}, and is propagated using the Trotterized circuit defined for timestep $\delta t$ for several total time intervals $T \in \{10, 100,1000\}$au. We consider several values of $N_{QL}$ for the number of QL-bits, comparing the full graph representation against the minimal (quotient) representation depending on computational feasibility. The main results of this section are as follows.

\begin{figure}
    \centering
    \includegraphics[width=\linewidth]{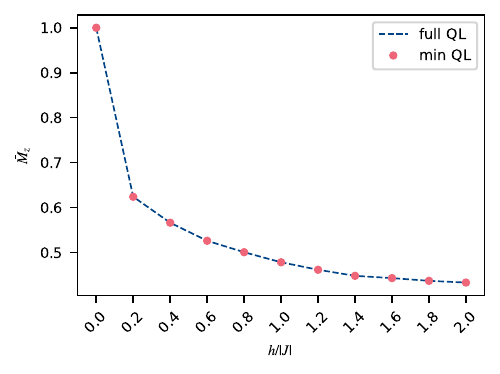}
    \caption{Comparing the time averaged magnetization as a function of transverse and coupling field ratio as simulated using the full graph Cartesian product and the minimal representation of the Cartesian product of two QL-bits. The minimal QL-bit quotient representation faithfully reproduces the full graph dynamics for $\bar{M}_{z}$.
    }
    \label{fig:fullvsmin}
\end{figure}

As stated in Section~\ref{subsec:states}, the full representation of the Cartesian product of $N_{QL}$ QL-bits scales as $(2n)^{N_{QL}}$, and Ref.~\cite{horvat_minimal_2026} constructs a minimal representation that brings down the resource scaling to $2^{N_{QL}}$ while preserving the  emergent eigenspectrum. 
To check how well the minimal representation captures the physics of the system relative to the full Cartesian product, we simulate the dynamics under both representations and compare the resulting $\bar{M}_{z}$ as a function of $h/|J|$ in both scenarios. \Cref{fig:fullvsmin} shows for this comparison for $N_{QL} = 2$, where the two representations agree up to numerical precision. We conclude that the minimal representation faithfully reproduces the full graph representation's dynamics. For the remainder of this section, all results presented are obtained from dynamics simulated on the quotient graphs of the QL-bits.

We track the change of average magnetization over time as a function of the coupling $J$ and transverse magnetic field strength $h$. The resulting quench dynamics can be divided into three regions. In the weak-field region, $h/|J| \ll 1$, the transverse field is a weak perturbation on the strong ferromagnetic coupling in the $z$-direction. The initial all-up state is close to the eigenstate of the post-quench Hamiltonian and therefore precesses only locally around the ordered configuration, and $\bar{M}_{z}$ stays close to $1$. In the strong-field $h/|J| \gg 1$ region, the transverse field dominates, driving the precession about the $x$-axis and largely overriding the nearest neighbor coupling. The coherent $z$-alignment washes out, and $\bar{M}_{z} \rightarrow 0 $ over time. Between these regions, $\bar{M}_{z}$ shows a rapid suppression near $h_{c} = h/|J| = 1$, consistent with the equilibrium critical point of the TFI model. This marks the transition from a dynamical regime where the initial ferromagnetic order is preserved to one where it is rapidly destroyed by the transverse field, and the QL-bit formalism captures this drop quantitatively, validating the circuit construction. Physically, the crossover coincides with $h_{c}$ because the same competition between the ferromagnetic coupling and the transverse field controls both the equilibrium order parameter and the post-quench magnetization.

\begin{figure}
    \centering
    \includegraphics[width=\linewidth]{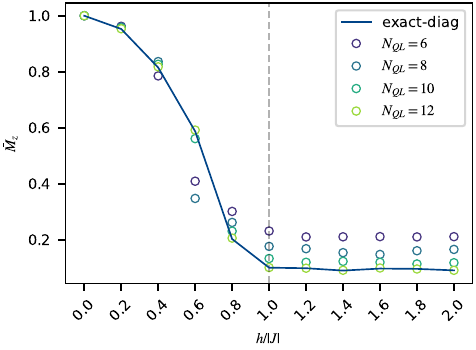}
    \caption{As the number of QL-bits $N_{QL}$ increases, the long-time average magnetization $\bar{M}_{z}$ converges to the exact curve. Exact diagonalization is computed for $N_{QL} = 12$ using eigenbasis propagation for the quench Hamiltonian, $\delta t = 0.1$, $T=100$au.}
    \label{fig:NQLconvergence}
\end{figure}

For comparison, we perform a classical simulation of the dynamics for $12$ spins using exact diagonalization and the eigenbasis propagation method.  Figure~\ref{fig:NQLconvergence} shows that as the number of QL-bits $N_{QL}$ increases, the long-time average magnetization $\bar{M}_{z}$ converges to this exact result. The accessible correlation length is limited by the system size, which results in a smooth transition instead of a sharp one near the critical point. This finite size effect diminishes as $N_{QL}$ increases. 
A sharp transition emerges only in the thermodynamic limit, which lies beyond the system sizes computationally accessible in the present implementation.

\begin{figure}
    \centering
    \includegraphics[width=\linewidth]{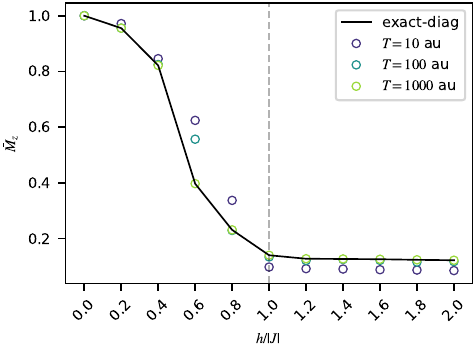}
    \caption{As the time averaging window increases, $\bar{M}_{z}$ converges to the exact curve. Exact diagonalization computed for $N_{QL} = 10$ using eigenbasis propagation for the quench Hamiltonian.}
    \label{fig:Tconvergence}
\end{figure}

The convergence of $\bar{M}_{z}$ with $T$, the time averaging window, is not uniform across $h/|J|$ as seen in \Cref{fig:Tconvergence}. At regions away from the critical region the short-window averages already agree closely with the exact result, since the post-quench oscillations complete multiple cycles within even $T=10$au. Near $h/|J|\approx0.6-0.8$, however, shorter times visibly overshoot the exact curve. As $h/|J|\rightarrow h_{c}$, the characteristic oscillation period of the post-quench dynamics grows, so a short averaging window captures only a partial cycle, biasing $\bar{M}_{z}$ toward its pre-quench value. Longer averaging windows resolve full oscillation periods even near $h_{c}$ converging to the exact diagonalization values. Note that exact diagonalization is performed for the same system size ($N_{QL}=10$) used in the QL-circuit simulation, isolating the effect of the averaging window $T$ from finite-size effects.

Our simulation results only show a qualitative agreement with the expected results and we expect better convergence with higher number of QL-bits. We do not claim to observe any new physics, but use the TFI model of spins to show how the QL-bit formalism can be used as an alternative to simulate quantum systems of such scale.

\section{Adjacency-Matrix Dynamics}\label{sec:adj_dyn}
The preceding results describe the dynamics of the encoded quantum state. We now change perspective and ask what those same unitary operations do to the graph adjacency matrix, $\mathcal{R}_{t} = U_{t} \mathcal{R}_{0} U_{t}^{\dagger}$. The synchronized compression $Q_t=P^\dagger \mathcal{R}_t P$ is a projection onto a subspace of $\mathcal{R}_t$ that preserves the emergent state representing the quantum system. The full adjacency matrix contains many additional node-to-node weights, however, and these microscopic weights can undergo substantial dynamics even when their net contribution to the synchronized compression, $Q_t$, is zero. We begin with an intuitive picture of this hidden motion before examining it more quantitatively.

\subsection{An Intuitive Picture of Microscopic Weight}

\begin{figure}[t!]
    \centering
    \includegraphics[width=\linewidth]{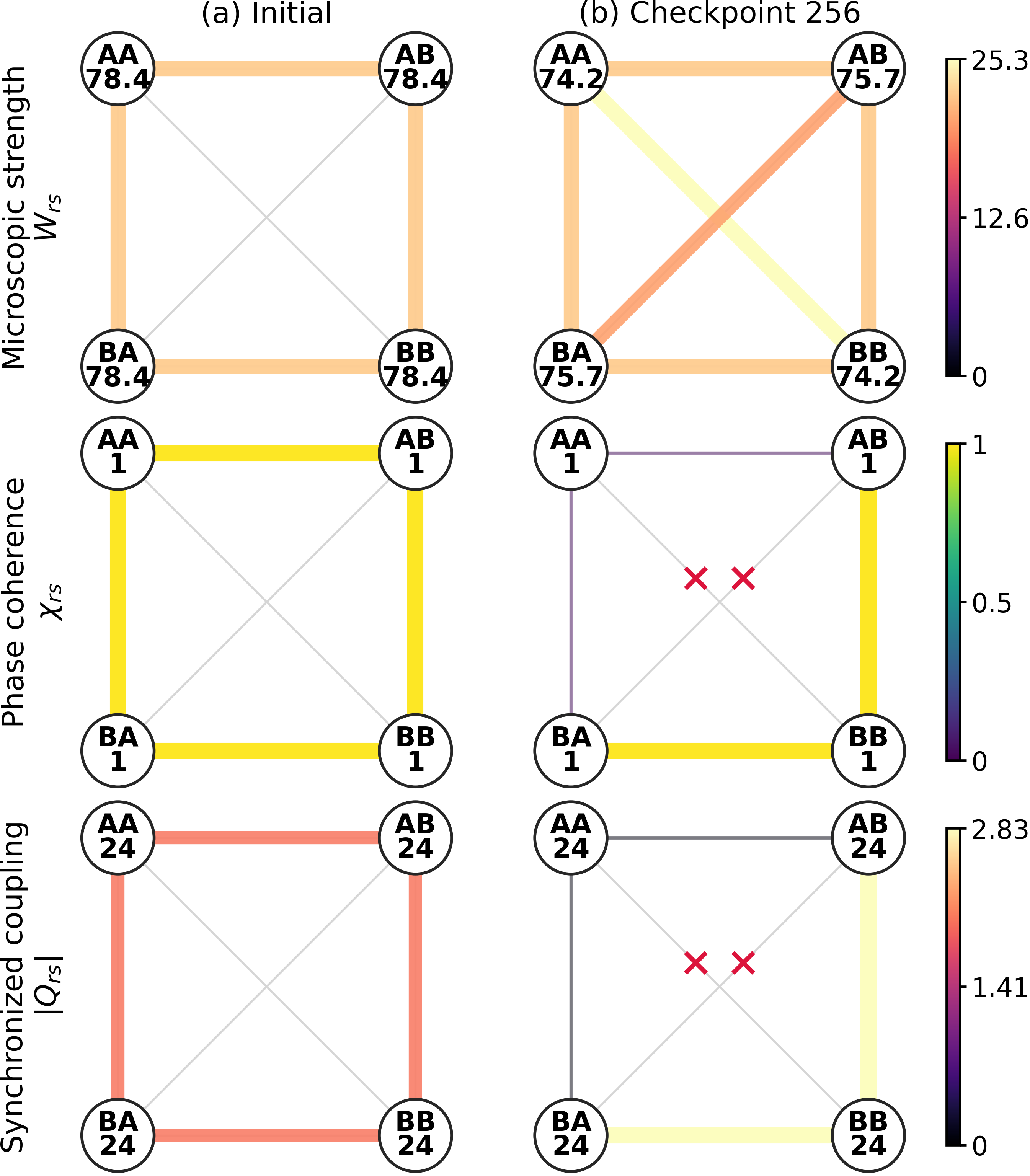}
    \caption{Metric-weighted four-region summaries of the two-QL-bit Cartesian product graph at \textbf{(a)} the initial checkpoint and \textbf{(b)} checkpoint $256$. The rows show microscopic block strength $W_{rs}$, weighted phase coherence $\chi_{rs}$, and synchronized coupling magnitude $|Q_{rs}|$, using one scale across both checkpoints within each row. Printed node values correspond to internal blocks, while both edge width and color encode inter-block values, with wider edges indicating larger values and the color bar showing the numerical scale. A red $\times$ marks a microscopically nonempty block whose displayed metric is numerically zero. At checkpoint $256$, substantial microscopic strength has developed in both cross-diagonal blocks, but phase cancellation leaves them absent from the synchronized compression. 
    }
    \label{fig:cartesian_block_metric_main}
\end{figure}

We illustrate these dynamics using the two-QL-bit graph, with $n=16$ vertices per subgraph, at the initial Trotter checkpoint and checkpoint $256$. Directly visualizing the full Cartesian product graph with $1024$ nodes at each checkpoint would obscure its large scale structure, so \Cref{fig:cartesian_block_metric_main} replaces the four Cartesian blocks $AA$, $AB$, $BA$, and $BB$ by four nodes. Each row then summarizes the full graph using one of the three diagnostics introduced in Section~\ref{sec:metrics}. The metric values at each node, therefore, describe the intra-block values (for example, $W_{AA-AA}$, $\chi_{AA-AA}$, and $Q_{AA-AA}$), while the width and color of the edges describe the inter-block connections. The bottom row is obtained from the synchronized compression; the upper two rows are block-wise summaries of microscopic strength and phase coherence.

The top row of Figure~\ref{fig:cartesian_block_metric_main} shows how the microscopic block strength redistributes over time for the different blocks of the Cartesian product graph. At checkpoint $0$, the inter-block edges lie entirely along the boundary of the Cartesian square, and the cross-diagonal blocks $AA-BB$ and $AB-BA$ are empty. By checkpoint $256$, a substantial number of edges have appeared between blocks that were initially decoupled ($AA-BB$, and $AB-BA$) while the number of edges within each block has decreased. Since the Frobenius norm of the full matrix is conserved by the unitary evolution, the natural visual picture is that the Trotter circuit continually redistributes edge weights throughout the full graph. This suggests an ``information mass'' picture, in which $W_{s}^2$, the squared microscopic weights, form a conserved total that is redistributed among the blocks, loosely analogous to operator weight descriptions of quantum information spreading \cite{vonkeyserlingk_operator_2018,xu_scrambling_2024}.

The middle row asks whether the phases corresponding to the complex edge weights within each block coherently align. The newly populated cross-diagonal blocks have weighted phase coherence $\chi_{rs}$ equal to numerical zero, represented by the red crosses in \Cref{fig:cartesian_block_metric_main}. These complex edge weights therefore carry substantial magnitude, but their phases interfere destructively when added. The consequence of this cancellation shows up in $|Q_{rs}|$, as shown in the bottom row: the effective synchronized couplings for the initially decoupled blocks, $|Q_{AA,BB}|$ and $|Q_{AB,BA}|$, remain at numerical zero even though the corresponding microscopic strengths are large.

The internal blocks show the complementary behavior. Their microscopic strengths change as mass is redistributed through the graph, while their phase coherence remains equal to one and their synchronized self-couplings remain fixed. The full graph can therefore rearrange a large amount of microscopic weight without changing what the synchronized modes see.

Taken together, the three rows provide a simple picture of the evolution: the Trotter circuit redistributes edge weights throughout the full graph, including new cross-diagonal connections, but the phases may partially or completely cancel when projected onto the synchronized subspace. The encoded state consequently responds only to the coherent portion that survives this cancellation. The extended comparison in the SI Appendix 
adds another checkpoint in time, showing that the edge weights continue to move between the two cross-diagonal channels while both remain canceled in the synchronized compression. Over the full $1000$-checkpoint trajectory, the cross-diagonal blocks repeatedly acquire substantial microscopic strength while their synchronized couplings remain zero to numerical precision (SI Appendix, Section~3A).

\subsection{Possible roles of the canceled cross-diagonal blocks}
The new cross-diagonal weight is part of an isospectral rearrangement rather than a change in the spectrum: $\mathcal{R}_t=U_t \mathcal{R}_0U_t^\dagger$ preserves both the spectrum and Frobenius norm. A plausible structural role is that these blocks carry microscopic or transverse operator weight while the synchronized compression remains sparse. This separation parallels equitable partition descriptions in which quotient dynamics occupy an invariant synchronized subspace while additional modes remain in the full network \cite{Schaub2016GraphPartitions}. The large-$W_{rs}$, zero-$\chi_{rs}$ pattern is also reminiscent of destructive-interference mechanisms in hopping networks, where nonzero microscopic paths cancel in a selected channel \cite{Vidal1998AharonovBohmCages}. This analogy does not, by itself, establish localization here.

\paragraph*{Possible connections to network neuroscience}
The distinction between microscopic connectivity and coupling between collective modes may be relevant to network neuroscience, where relating network structure to collective activity is a central question \cite{bassett_network_2017}. In our setting, the additional cross-diagonal connections suggest a possible structural ``cut through the middle'' interpretation. At checkpoint 0, the unweighted four-region support graph is the boundary cycle with edges $AA-AB$, $AB-BB$, $BB-BA$, and $BA-AA$. When both cross-diagonal blocks become nonempty while the boundary connections remain present, this support becomes the complete four-region graph. Within this representation, the distance between each opposite pair drops from 2 to 1, and the mean distance over the six unordered region pairs drops from $4/3$ to 1. This resembles the shortcuts studied in small-world networks \cite{cirino_connectivity_2026,rodrigues_kuramoto_2016}, although its significance for communication in the present setting remains open. In particular, these distances describe connectivity between regions without accounting for coupling magnitudes or phases, so their reduction alone does not establish faster coherent transfer.

In a percolated hierarchical small-world network, increasing connectivity can eventually reduce root-to-boundary transport because the additional links reorganize spectral pathways and promote interference-induced recirculation \cite{cirino_connectivity_2026}. The comparison is necessarily qualitative: bond percolation changes the transport Hamiltonian in that setting, whereas $\mathcal{R}_t$ here is related to $\mathcal{R}_0$ by a unitary similarity transformation and therefore has the same spectrum. The result therefore supplies a caution rather than a mechanism for the present dynamics. Because the cross-diagonal blocks cancel in $Q_t$, our data establish new microscopic support but not faster transport of the synchronized state. Resolving their transport role would require a specified phase-sensitive propagator, current, communicability, or ablation observable.
The distinct $AA-BB$ and $AB-BA$ frequencies nevertheless indicate different patterns of microscopic reorganization in the two cross-diagonal blocks (SI Appendix, Section~3B).

Altogether, these results separate the synchronized dynamics retained by the quotient representation from the microscopic reorganization of the full graph. A direct comparison confirms that the full Cartesian product and contracted calculations reproduce the same encoded-state trajectory and synchronized operator evolution to numerical precision, while the full graph develops phase-canceled cross-diagonal blocks with pair-dependent recurrence frequencies that are absent from the quotient description. The contraction thus preserves the synchronized dynamics examined here, whereas the full representation is required to resolve the additional node-level structure.

\section{Outlook}\label{sec:outlook}
Our simulation results of Section~\ref{sec:results} illustrate that distinguishing a genuine phase transition from a smooth finite-size feature in principle requires accessing larger $N_{QL}$, so that the critical $h/|J|$ value converges to its thermodynamic limit. Scaling the QL-bit construction to larger $N_{QL}$ is computationally costly, however, since the adjacency matrix of the Cartesian product graph grows exponentially with the number of QL-bits. This finite-size scaling limitation is a property of the Cartesian product's computational cost, distinct from the question of physical implementations, as discussed below.

The graph diagnostics of Section~\ref{sec:adj_dyn} revealed that the unitary transformations of the QL-circuit generate complex edge weights on the graph adjacency matrix, as well as additional edges between non-nearest neighbors in the original Cartesian product graph. 
A physical realization of these results must therefore satisfy two requirements: long-range connections and the capacity to support complex-valued edges. Studies of synchronization in complex weighted networks, in particular, the Sakaguchi-Kuramoto model, offer a qualitative comparison in which phase frustrated coupling has been shown to shift the critical coupling strength required for spontaneous synchronization~\cite{rodrigues_kuramoto_2016}. Phase-lagged Kuramoto coupling does not, however, reproduce the same synchronization structure as a general complex Hermitian coupling matrix, so whether an analogous shift occurs for the couplings generated by our QL-circuits remains an open question.

The requirement for long-range connectivity is not merely an engineering inconvenience but reflects a more fundamental obstruction. Barzdin and Kolmogorov showed that realizing an expander graph in $\mathbb{R}^3$ space requires thick, separated connections, so that long-range connectivity must be paid for in spatial volume~\cite{kolmogorov_realization_1993}. Since the QL-bit construction is predicated on graphs with an expander property~\cite{amati_quantum_2025}, any spatially local physical implementation of a QL system faces an intrinsic tension between wiring cost and the connectivity needed to support the desired synchronized dynamics.

Potential physical platforms, therefore, differ in how they might navigate this tension. CMOS systems with coupled LC oscillators~\cite{csaba_coupled_2020} and spin-torque oscillator networks with long-range electrical coupling~\cite{lebrun_mutual_2017,tsunegi_scaling_2018} 
both offer routes to long-range coupling that partially sidestep the volume cost of local wiring. As an alternative, biological systems, whose networks exhibit heterogeneous expansion and regularity properties rather than a strict expander structure~\cite{kozachkov_neuronastrocyte_2025}, may suffice for the synchronized dynamics of interest. Moving from simulations of QL-systems to physical implementations will build on existing work that has implemented features of quantum mechanics in non-quantum mechanical networks~\cite{kaiser_probabilistic_2021}. Exploring the limitations of QL-based computation will help define bounds on how quantum information can be represented in classical networks, such as those discussed here.

\vspace{1mm}
\emph{Acknowledgments}---GDS thanks the Air Force Office of Scientific Research, award number FA9550261B154, for support of this work. SK thanks the U.S. Department of Energy (DOE), Office of Basic Energy Sciences, under Grant No.~DE-SC0026309.
\FloatBarrier
\bibliography{references/expandergraphs,references/generalnetworks,references/minimalrep,references/QL-ditsPaper,references/quantumcomputing,references/smallworldnetworks,references/refs}

\begin{thebibliography}{49}%
\makeatletter
\providecommand \@ifxundefined [1]{%
 \@ifx{#1\undefined}
}%
\providecommand \@ifnum [1]{%
 \ifnum #1\expandafter \@firstoftwo
 \else \expandafter \@secondoftwo
 \fi
}%
\providecommand \@ifx [1]{%
 \ifx #1\expandafter \@firstoftwo
 \else \expandafter \@secondoftwo
 \fi
}%
\providecommand \natexlab [1]{#1}%
\providecommand \enquote  [1]{``#1''}%
\providecommand \bibnamefont  [1]{#1}%
\providecommand \bibfnamefont [1]{#1}%
\providecommand \citenamefont [1]{#1}%
\providecommand \href@noop [0]{\@secondoftwo}%
\providecommand \href [0]{\begingroup \@sanitize@url \@href}%
\providecommand \@href[1]{\@@startlink{#1}\@@href}%
\providecommand \@@href[1]{\endgroup#1\@@endlink}%
\providecommand \@sanitize@url [0]{\catcode `\\12\catcode `\$12\catcode `\&12\catcode `\#12\catcode `\^12\catcode `\_12\catcode `\%12\relax}%
\providecommand \@@startlink[1]{}%
\providecommand \@@endlink[0]{}%
\providecommand \url  [0]{\begingroup\@sanitize@url \@url }%
\providecommand \@url [1]{\endgroup\@href {#1}{\urlprefix }}%
\providecommand \urlprefix  [0]{URL }%
\providecommand \Eprint [0]{\href }%
\providecommand \doibase [0]{https://doi.org/}%
\providecommand \selectlanguage [0]{\@gobble}%
\providecommand \bibinfo  [0]{\@secondoftwo}%
\providecommand \bibfield  [0]{\@secondoftwo}%
\providecommand \translation [1]{[#1]}%
\providecommand \BibitemOpen [0]{}%
\providecommand \bibitemStop [0]{}%
\providecommand \bibitemNoStop [0]{.\EOS\space}%
\providecommand \EOS [0]{\spacefactor3000\relax}%
\providecommand \BibitemShut  [1]{\csname bibitem#1\endcsname}%
\let\auto@bib@innerbib\@empty
\bibitem [{\citenamefont {Hopfield}(1982)}]{hopfield_neuralnets_1982}%
  \BibitemOpen
  \bibfield  {author} {\bibinfo {author} {\bibfnamefont {J.~J.}\ \bibnamefont {Hopfield}},\ }\bibfield  {title} {\bibinfo {title} {Neural networks and physical systems with emergent collective computational abilities.},\ }\href {https://doi.org/10.1073/pnas.79.8.2554} {\bibfield  {journal} {\bibinfo  {journal} {Proceedings of the National Academy of Sciences}\ }\textbf {\bibinfo {volume} {79}},\ \bibinfo {pages} {2554} (\bibinfo {year} {1982})},\ \Eprint {https://arxiv.org/abs/https://www.pnas.org/doi/pdf/10.1073/pnas.79.8.2554} {https://www.pnas.org/doi/pdf/10.1073/pnas.79.8.2554} \BibitemShut {NoStop}%
\bibitem [{\citenamefont {Bassett}\ and\ \citenamefont {Sporns}(2017)}]{bassett_network_2017}%
  \BibitemOpen
  \bibfield  {author} {\bibinfo {author} {\bibfnamefont {D.~S.}\ \bibnamefont {Bassett}}\ and\ \bibinfo {author} {\bibfnamefont {O.}~\bibnamefont {Sporns}},\ }\bibfield  {title} {\bibinfo {title} {Network neuroscience},\ }\href {https://doi.org/10.1038/nn.4502} {\bibfield  {journal} {\bibinfo  {journal} {Nature Neuroscience}\ }\textbf {\bibinfo {volume} {20}},\ \bibinfo {pages} {353} (\bibinfo {year} {2017})}\BibitemShut {NoStop}%
\bibitem [{\citenamefont {Vicsek}\ \emph {et~al.}(1995)\citenamefont {Vicsek}, \citenamefont {Czir\'{o}k}, \citenamefont {Ben-Jacob}, \citenamefont {Cohen},\ and\ \citenamefont {Shochet}}]{vicsek_novel_1995}%
  \BibitemOpen
  \bibfield  {author} {\bibinfo {author} {\bibfnamefont {T.}~\bibnamefont {Vicsek}}, \bibinfo {author} {\bibfnamefont {A.}~\bibnamefont {Czir\'{o}k}}, \bibinfo {author} {\bibfnamefont {E.}~\bibnamefont {Ben-Jacob}}, \bibinfo {author} {\bibfnamefont {I.}~\bibnamefont {Cohen}},\ and\ \bibinfo {author} {\bibfnamefont {O.}~\bibnamefont {Shochet}},\ }\bibfield  {title} {\bibinfo {title} {Novel type of phase transition in a system of self-driven particles},\ }\href {https://doi.org/10.1103/PhysRevLett.75.1226} {\bibfield  {journal} {\bibinfo  {journal} {Physical Review Letters}\ }\textbf {\bibinfo {volume} {75}},\ \bibinfo {pages} {1226} (\bibinfo {year} {1995})}\BibitemShut {NoStop}%
\bibitem [{\citenamefont {Pastor-Satorras}\ and\ \citenamefont {Vespignani}(2001)}]{pastor-satorras_epidemic_2001}%
  \BibitemOpen
  \bibfield  {author} {\bibinfo {author} {\bibfnamefont {R.}~\bibnamefont {Pastor-Satorras}}\ and\ \bibinfo {author} {\bibfnamefont {A.}~\bibnamefont {Vespignani}},\ }\bibfield  {title} {\bibinfo {title} {Epidemic spreading in scale-free networks},\ }\href {https://doi.org/10.1103/PhysRevLett.86.3200} {\bibfield  {journal} {\bibinfo  {journal} {Physical Review Letters}\ }\textbf {\bibinfo {volume} {86}},\ \bibinfo {pages} {3200} (\bibinfo {year} {2001})}\BibitemShut {NoStop}%
\bibitem [{\citenamefont {Csaba}\ and\ \citenamefont {Porod}(2020)}]{csaba_coupled_2020}%
  \BibitemOpen
  \bibfield  {author} {\bibinfo {author} {\bibfnamefont {G.}~\bibnamefont {Csaba}}\ and\ \bibinfo {author} {\bibfnamefont {W.}~\bibnamefont {Porod}},\ }\bibfield  {title} {\bibinfo {title} {Coupled oscillators for computing: A review and perspective},\ }\href {https://doi.org/10.1063/1.5120412} {\bibfield  {journal} {\bibinfo  {journal} {Applied Physics Reviews}\ }\textbf {\bibinfo {volume} {7}},\ \bibinfo {pages} {011302} (\bibinfo {year} {2020})}\BibitemShut {NoStop}%
\bibitem [{\citenamefont {Jaeger}\ and\ \citenamefont {Haas}(2004)}]{jaeger_harnessing_2004}%
  \BibitemOpen
  \bibfield  {author} {\bibinfo {author} {\bibfnamefont {H.}~\bibnamefont {Jaeger}}\ and\ \bibinfo {author} {\bibfnamefont {H.}~\bibnamefont {Haas}},\ }\bibfield  {title} {\bibinfo {title} {Harnessing nonlinearity: Predicting chaotic systems and saving energy in wireless communication},\ }\href@noop {} {\bibfield  {journal} {\bibinfo  {journal} {science}\ }\textbf {\bibinfo {volume} {304}},\ \bibinfo {pages} {78} (\bibinfo {year} {2004})}\BibitemShut {NoStop}%
\bibitem [{\citenamefont {Traversa}\ \emph {et~al.}(2014)\citenamefont {Traversa}, \citenamefont {Bonani}, \citenamefont {Pershin},\ and\ \citenamefont {Di~Ventra}}]{traversa_dcram_2014}%
  \BibitemOpen
  \bibfield  {author} {\bibinfo {author} {\bibfnamefont {F.~L.}\ \bibnamefont {Traversa}}, \bibinfo {author} {\bibfnamefont {F.}~\bibnamefont {Bonani}}, \bibinfo {author} {\bibfnamefont {Y.~V.}\ \bibnamefont {Pershin}},\ and\ \bibinfo {author} {\bibfnamefont {M.}~\bibnamefont {Di~Ventra}},\ }\bibfield  {title} {\bibinfo {title} {Dynamic computing random access memory},\ }\href {https://doi.org/10.1088/0957-4484/25/28/285201} {\bibfield  {journal} {\bibinfo  {journal} {Nanotechnology}\ }\textbf {\bibinfo {volume} {25}},\ \bibinfo {pages} {285201} (\bibinfo {year} {2014})}\BibitemShut {NoStop}%
\bibitem [{\citenamefont {Mead}(1990)}]{mead_neuromorphic_1990}%
  \BibitemOpen
  \bibfield  {author} {\bibinfo {author} {\bibfnamefont {C.}~\bibnamefont {Mead}},\ }\bibfield  {title} {\bibinfo {title} {Neuromorphic electronic systems},\ }\href@noop {} {\bibfield  {journal} {\bibinfo  {journal} {Proceedings of the IEEE}\ }\textbf {\bibinfo {volume} {78}},\ \bibinfo {pages} {1629} (\bibinfo {year} {1990})}\BibitemShut {NoStop}%
\bibitem [{\citenamefont {Scholes}(2024)}]{scholes_quantumlike_2024}%
  \BibitemOpen
  \bibfield  {author} {\bibinfo {author} {\bibfnamefont {G.~D.}\ \bibnamefont {Scholes}},\ }\bibfield  {title} {\bibinfo {title} {Quantum-like states on complex synchronized networks},\ }\href {https://doi.org/10.1098/rspa.2024.0209} {\bibfield  {journal} {\bibinfo  {journal} {Proceedings of the Royal Society A: Mathematical, Physical and Engineering Sciences}\ }\textbf {\bibinfo {volume} {480}},\ \bibinfo {pages} {20240209} (\bibinfo {year} {2024})}\BibitemShut {NoStop}%
\bibitem [{\citenamefont {Scholes}\ and\ \citenamefont {Amati}(2025)}]{scholes_qlcartpdt_2025}%
  \BibitemOpen
  \bibfield  {author} {\bibinfo {author} {\bibfnamefont {G.~D.}\ \bibnamefont {Scholes}}\ and\ \bibinfo {author} {\bibfnamefont {G.}~\bibnamefont {Amati}},\ }\bibfield  {title} {\bibinfo {title} {Quantum-like {Product} {States} {Constructed} from {Classical} {Networks}},\ }\href {https://doi.org/10.1103/PhysRevLett.134.060202} {\bibfield  {journal} {\bibinfo  {journal} {Physical Review Letters}\ }\textbf {\bibinfo {volume} {134}},\ \bibinfo {pages} {060202} (\bibinfo {year} {2025})}\BibitemShut {NoStop}%
\bibitem [{\citenamefont {Amati}\ and\ \citenamefont {Scholes}(2025)}]{amati_quantum_2025}%
  \BibitemOpen
  \bibfield  {author} {\bibinfo {author} {\bibfnamefont {G.}~\bibnamefont {Amati}}\ and\ \bibinfo {author} {\bibfnamefont {G.~D.}\ \bibnamefont {Scholes}},\ }\bibfield  {title} {\bibinfo {title} {Quantum information with quantumlike bits},\ }\href {https://doi.org/10.1103/PhysRevA.111.062203} {\bibfield  {journal} {\bibinfo  {journal} {Physical Review A}\ }\textbf {\bibinfo {volume} {111}},\ \bibinfo {pages} {062203} (\bibinfo {year} {2025})}\BibitemShut {NoStop}%
\bibitem [{\citenamefont {Verstraete}\ \emph {et~al.}(2008)\citenamefont {Verstraete}, \citenamefont {Murg},\ and\ \citenamefont {Cirac}}]{verstraete_matrix_2008}%
  \BibitemOpen
  \bibfield  {author} {\bibinfo {author} {\bibfnamefont {F.}~\bibnamefont {Verstraete}}, \bibinfo {author} {\bibfnamefont {V.}~\bibnamefont {Murg}},\ and\ \bibinfo {author} {\bibfnamefont {J.~I.}\ \bibnamefont {Cirac}},\ }\bibfield  {title} {\bibinfo {title} {Matrix product states, projected entangled pair states, and variational renormalization group methods for quantum spin systems},\ }\href {https://doi.org/10.1080/14789940801912366} {\bibfield  {journal} {\bibinfo  {journal} {Adv. Phys.}\ }\textbf {\bibinfo {volume} {57}},\ \bibinfo {pages} {143} (\bibinfo {year} {2008})}\BibitemShut {NoStop}%
\bibitem [{\citenamefont {Orús}(2014)}]{orus_mpspeps_2014}%
  \BibitemOpen
  \bibfield  {author} {\bibinfo {author} {\bibfnamefont {R.}~\bibnamefont {Orús}},\ }\bibfield  {title} {\bibinfo {title} {A practical introduction to tensor networks: Matrix product states and projected entangled pair states},\ }\href {https://doi.org/https://doi.org/10.1016/j.aop.2014.06.013} {\bibfield  {journal} {\bibinfo  {journal} {Annals of Physics}\ }\textbf {\bibinfo {volume} {349}},\ \bibinfo {pages} {117} (\bibinfo {year} {2014})}\BibitemShut {NoStop}%
\bibitem [{\citenamefont {Fauseweh}(2024)}]{fauseweh_quantum_2024}%
  \BibitemOpen
  \bibfield  {author} {\bibinfo {author} {\bibfnamefont {B.}~\bibnamefont {Fauseweh}},\ }\bibfield  {title} {\bibinfo {title} {Quantum many-body simulations on digital quantum computers: State-of-the-art and future challenges},\ }\href {https://doi.org/10.1038/s41467-024-46402-9} {\bibfield  {journal} {\bibinfo  {journal} {Nature Communications}\ }\textbf {\bibinfo {volume} {15}},\ \bibinfo {pages} {2123} (\bibinfo {year} {2024})}\BibitemShut {NoStop}%
\bibitem [{\citenamefont {Kolmogorov}\ and\ \citenamefont {Barzdin}(1993)}]{kolmogorov_realization_1993}%
  \BibitemOpen
  \bibfield  {author} {\bibinfo {author} {\bibfnamefont {A.~N.}\ \bibnamefont {Kolmogorov}}\ and\ \bibinfo {author} {\bibfnamefont {Y.~M.}\ \bibnamefont {Barzdin}},\ }\bibfield  {title} {\bibinfo {title} {On the realization of networks in three-dimensional space},\ }in\ \href@noop {} {\emph {\bibinfo {booktitle} {Selected Works of A. N. Kolmogorov, Volume 3: Information Theory and the Theory of Algorithms}}},\ \bibinfo {editor} {edited by\ \bibinfo {editor} {\bibfnamefont {A.~N.}\ \bibnamefont {Shiryaev}}}\ (\bibinfo  {publisher} {Kluwer Academic Publishers},\ \bibinfo {address} {Dordrecht},\ \bibinfo {year} {1993})\ pp.\ \bibinfo {pages} {194--202},\ \bibinfo {note} {english translation of the 1967 Russian paper published in \emph{Problemy Kibernetiki}, Vol. 19, pp. 261--268}\BibitemShut {NoStop}%
\bibitem [{\citenamefont {Dickey}\ \emph {et~al.}(2026)\citenamefont {Dickey}, \citenamefont {Vyas},\ and\ \citenamefont {Kais}}]{dickey_programmable_2026}%
  \BibitemOpen
  \bibfield  {author} {\bibinfo {author} {\bibfnamefont {E.}~\bibnamefont {Dickey}}, \bibinfo {author} {\bibfnamefont {A.}~\bibnamefont {Vyas}},\ and\ \bibinfo {author} {\bibfnamefont {S.}~\bibnamefont {Kais}},\ }\bibfield  {title} {\bibinfo {title} {Programmable quantum-like bits from signed regular graphs},\ }\href {https://doi.org/10.1098/rsos.260238} {\bibfield  {journal} {\bibinfo  {journal} {Royal Society Open Science}\ }\textbf {\bibinfo {volume} {13}},\ \bibinfo {pages} {rsos260238} (\bibinfo {year} {2026})},\ \Eprint {https://arxiv.org/abs/https://royalsocietypublishing.org/rsos/article-pdf/doi/10.1098/rsos.260238/6166352/rsos.260238.pdf} {https://royalsocietypublishing.org/rsos/article-pdf/doi/10.1098/rsos.260238/6166352/rsos.260238.pdf} \BibitemShut {NoStop}%
\bibitem [{\citenamefont {Dickey}\ and\ \citenamefont {Kais}(2026)}]{dickey2026complex}%
  \BibitemOpen
  \bibfield  {author} {\bibinfo {author} {\bibfnamefont {E.}~\bibnamefont {Dickey}}\ and\ \bibinfo {author} {\bibfnamefont {S.}~\bibnamefont {Kais}},\ }\href {https://arxiv.org/abs/2604.23991} {\bibinfo {title} {Universal complex {Quantum-Like} bits from {Hermitian} weighted graphs}},\ \bibinfo {howpublished} {arXiv:2604.23991} (\bibinfo {year} {2026}),\ \bibinfo {note} {under review},\ \Eprint {https://arxiv.org/abs/2604.23991} {arXiv:2604.23991 [quant-ph]} \BibitemShut {NoStop}%
\bibitem [{\citenamefont {Saha}\ and\ \citenamefont {Scholes}(2025)}]{saha_multi-dimensional_2025}%
  \BibitemOpen
  \bibfield  {author} {\bibinfo {author} {\bibfnamefont {D.}~\bibnamefont {Saha}}\ and\ \bibinfo {author} {\bibfnamefont {G.~D.}\ \bibnamefont {Scholes}},\ }\bibfield  {title} {\bibinfo {title} {Multi-{Dimensional} {Quantum}-like {Resources} from {Complex} {Synchronized} {Networks}},\ }\href {https://doi.org/10.3390/e27090963} {\bibfield  {journal} {\bibinfo  {journal} {Entropy}\ }\textbf {\bibinfo {volume} {27}},\ \bibinfo {pages} {963} (\bibinfo {year} {2025})}\BibitemShut {NoStop}%
\bibitem [{\citenamefont {Xie}\ \emph {et~al.}(2026)\citenamefont {Xie}, \citenamefont {Dickey}, \citenamefont {Taimoor},\ and\ \citenamefont {Kais}}]{xie_dynamicalreadout_2026}%
  \BibitemOpen
  \bibfield  {author} {\bibinfo {author} {\bibfnamefont {S.}~\bibnamefont {Xie}}, \bibinfo {author} {\bibfnamefont {E.}~\bibnamefont {Dickey}}, \bibinfo {author} {\bibfnamefont {S.~A.}\ \bibnamefont {Taimoor}},\ and\ \bibinfo {author} {\bibfnamefont {S.}~\bibnamefont {Kais}},\ }\href {https://arxiv.org/abs/2609.14620} {\bibinfo {title} {Dynamical readout of measurement statistics and emergent entanglement-like states in classical networks}} (\bibinfo {year} {2026}),\ \Eprint {https://arxiv.org/abs/2609.14620} {arXiv:2609.14620 [quant-ph]} \BibitemShut {NoStop}%
\bibitem [{\citenamefont {Horvat}\ \emph {et~al.}(2026)\citenamefont {Horvat}, \citenamefont {Narang}, \citenamefont {Saha},\ and\ \citenamefont {Scholes}}]{horvat_minimal_2026}%
  \BibitemOpen
  \bibfield  {author} {\bibinfo {author} {\bibfnamefont {W.}~\bibnamefont {Horvat}}, \bibinfo {author} {\bibfnamefont {P.}~\bibnamefont {Narang}}, \bibinfo {author} {\bibfnamefont {D.}~\bibnamefont {Saha}},\ and\ \bibinfo {author} {\bibfnamefont {G.~D.}\ \bibnamefont {Scholes}},\ }\bibfield  {title} {\bibinfo {title} {Minimal representations of topology-preserving quantumlike states},\ }\href {https://doi.org/10.1103/4kcg-z9v2} {\bibfield  {journal} {\bibinfo  {journal} {Phys. Rev. E}\ } (\bibinfo {year} {2026})}\BibitemShut {NoStop}%
\bibitem [{\citenamefont {Barenco}\ \emph {et~al.}(1995)\citenamefont {Barenco}, \citenamefont {Bennett}, \citenamefont {Cleve}, \citenamefont {DiVincenzo}, \citenamefont {Margolus}, \citenamefont {Shor}, \citenamefont {Sleator}, \citenamefont {Smolin},\ and\ \citenamefont {Weinfurter}}]{barenco_elementary_1995}%
  \BibitemOpen
  \bibfield  {author} {\bibinfo {author} {\bibfnamefont {A.}~\bibnamefont {Barenco}}, \bibinfo {author} {\bibfnamefont {C.~H.}\ \bibnamefont {Bennett}}, \bibinfo {author} {\bibfnamefont {R.}~\bibnamefont {Cleve}}, \bibinfo {author} {\bibfnamefont {D.~P.}\ \bibnamefont {DiVincenzo}}, \bibinfo {author} {\bibfnamefont {N.}~\bibnamefont {Margolus}}, \bibinfo {author} {\bibfnamefont {P.}~\bibnamefont {Shor}}, \bibinfo {author} {\bibfnamefont {T.}~\bibnamefont {Sleator}}, \bibinfo {author} {\bibfnamefont {J.~A.}\ \bibnamefont {Smolin}},\ and\ \bibinfo {author} {\bibfnamefont {H.}~\bibnamefont {Weinfurter}},\ }\bibfield  {title} {\bibinfo {title} {Elementary gates for quantum computation},\ }\href {https://doi.org/10.1103/PhysRevA.52.3457} {\bibfield  {journal} {\bibinfo  {journal} {Phys. Rev. A}\ }\textbf {\bibinfo {volume} {52}},\ \bibinfo {pages} {3457} (\bibinfo {year} {1995})}\BibitemShut {NoStop}%
\bibitem [{\citenamefont {Pfeuty}(1970)}]{pfeuty_tfi1D_1970}%
  \BibitemOpen
  \bibfield  {author} {\bibinfo {author} {\bibfnamefont {P.}~\bibnamefont {Pfeuty}},\ }\bibfield  {title} {\bibinfo {title} {The one-dimensional ising model with a transverse field},\ }\href@noop {} {\bibfield  {journal} {\bibinfo  {journal} {ANNALS of Physics}\ }\textbf {\bibinfo {volume} {57}},\ \bibinfo {pages} {79} (\bibinfo {year} {1970})}\BibitemShut {NoStop}%
\bibitem [{\citenamefont {Sachdev}(2011)}]{sachdev_quantum_2011}%
  \BibitemOpen
  \bibfield  {author} {\bibinfo {author} {\bibfnamefont {S.}~\bibnamefont {Sachdev}},\ }\href {https://doi.org/10.1017/CBO9780511973765} {\emph {\bibinfo {title} {Quantum Phase Transitions}}},\ \bibinfo {edition} {2nd}\ ed.\ (\bibinfo  {publisher} {Cambridge University Press},\ \bibinfo {address} {Cambridge},\ \bibinfo {year} {2011})\BibitemShut {NoStop}%
\bibitem [{\citenamefont {Heyl}\ \emph {et~al.}(2013)\citenamefont {Heyl}, \citenamefont {Polkovnikov},\ and\ \citenamefont {Kehrein}}]{heyl_dqpt_2013}%
  \BibitemOpen
  \bibfield  {author} {\bibinfo {author} {\bibfnamefont {M.}~\bibnamefont {Heyl}}, \bibinfo {author} {\bibfnamefont {A.}~\bibnamefont {Polkovnikov}},\ and\ \bibinfo {author} {\bibfnamefont {S.}~\bibnamefont {Kehrein}},\ }\bibfield  {title} {\bibinfo {title} {Dynamical quantum phase transitions in the transverse-field ising model},\ }\href {https://doi.org/10.1103/PhysRevLett.110.135704} {\bibfield  {journal} {\bibinfo  {journal} {Phys. Rev. Lett.}\ }\textbf {\bibinfo {volume} {110}},\ \bibinfo {pages} {135704} (\bibinfo {year} {2013})}\BibitemShut {NoStop}%
\bibitem [{\citenamefont {Kim}\ \emph {et~al.}(2010)\citenamefont {Kim}, \citenamefont {Chang}, \citenamefont {Korenblit}, \citenamefont {Islam}, \citenamefont {Edwards}, \citenamefont {Freericks}, \citenamefont {Lin}, \citenamefont {Duan},\ and\ \citenamefont {Monroe}}]{kim_quantum_2010}%
  \BibitemOpen
  \bibfield  {author} {\bibinfo {author} {\bibfnamefont {K.}~\bibnamefont {Kim}}, \bibinfo {author} {\bibfnamefont {M.-S.}\ \bibnamefont {Chang}}, \bibinfo {author} {\bibfnamefont {S.}~\bibnamefont {Korenblit}}, \bibinfo {author} {\bibfnamefont {R.}~\bibnamefont {Islam}}, \bibinfo {author} {\bibfnamefont {E.~E.}\ \bibnamefont {Edwards}}, \bibinfo {author} {\bibfnamefont {J.~K.}\ \bibnamefont {Freericks}}, \bibinfo {author} {\bibfnamefont {G.-D.}\ \bibnamefont {Lin}}, \bibinfo {author} {\bibfnamefont {L.-M.}\ \bibnamefont {Duan}},\ and\ \bibinfo {author} {\bibfnamefont {C.}~\bibnamefont {Monroe}},\ }\bibfield  {title} {\bibinfo {title} {Quantum simulation of frustrated ising spins with trapped ions},\ }\href@noop {} {\bibfield  {journal} {\bibinfo  {journal} {Nature}\ }\textbf {\bibinfo {volume} {465}},\ \bibinfo {pages} {590} (\bibinfo {year} {2010})}\BibitemShut {NoStop}%
\bibitem [{\citenamefont {Bernien}\ \emph {et~al.}(2017)\citenamefont {Bernien}, \citenamefont {Schwartz}, \citenamefont {Keesling}, \citenamefont {Levine}, \citenamefont {Omran}, \citenamefont {Pichler}, \citenamefont {Choi}, \citenamefont {Zibrov}, \citenamefont {Endres}, \citenamefont {Greiner}, \citenamefont {Vuletić},\ and\ \citenamefont {Lukin}}]{bernien_probing_2017}%
  \BibitemOpen
  \bibfield  {author} {\bibinfo {author} {\bibfnamefont {H.}~\bibnamefont {Bernien}}, \bibinfo {author} {\bibfnamefont {S.}~\bibnamefont {Schwartz}}, \bibinfo {author} {\bibfnamefont {A.}~\bibnamefont {Keesling}}, \bibinfo {author} {\bibfnamefont {H.}~\bibnamefont {Levine}}, \bibinfo {author} {\bibfnamefont {A.}~\bibnamefont {Omran}}, \bibinfo {author} {\bibfnamefont {H.}~\bibnamefont {Pichler}}, \bibinfo {author} {\bibfnamefont {S.}~\bibnamefont {Choi}}, \bibinfo {author} {\bibfnamefont {A.~S.}\ \bibnamefont {Zibrov}}, \bibinfo {author} {\bibfnamefont {M.}~\bibnamefont {Endres}}, \bibinfo {author} {\bibfnamefont {M.}~\bibnamefont {Greiner}}, \bibinfo {author} {\bibfnamefont {V.}~\bibnamefont {Vuletić}},\ and\ \bibinfo {author} {\bibfnamefont {M.~D.}\ \bibnamefont {Lukin}},\ }\bibfield  {title} {\bibinfo {title} {Probing many-body dynamics on a 51-atom quantum simulator},\ }\href@noop {} {\bibfield  {journal} {\bibinfo  {journal} {Nature}\ }\textbf {\bibinfo {volume} {551}},\ \bibinfo {pages} {579} (\bibinfo
  {year} {2017})}\BibitemShut {NoStop}%
\bibitem [{\citenamefont {Dborin}\ \emph {et~al.}(2022)\citenamefont {Dborin}, \citenamefont {Wimalaweera}, \citenamefont {Barratt}, \citenamefont {Ostby}, \citenamefont {O’Brien},\ and\ \citenamefont {Green}}]{dborin_simulating_2022}%
  \BibitemOpen
  \bibfield  {author} {\bibinfo {author} {\bibfnamefont {J.}~\bibnamefont {Dborin}}, \bibinfo {author} {\bibfnamefont {V.}~\bibnamefont {Wimalaweera}}, \bibinfo {author} {\bibfnamefont {F.}~\bibnamefont {Barratt}}, \bibinfo {author} {\bibfnamefont {E.}~\bibnamefont {Ostby}}, \bibinfo {author} {\bibfnamefont {T.~E.}\ \bibnamefont {O’Brien}},\ and\ \bibinfo {author} {\bibfnamefont {A.~G.}\ \bibnamefont {Green}},\ }\bibfield  {title} {\bibinfo {title} {Simulating groundstate and dynamical quantum phase transitions on a superconducting quantum computer},\ }\href@noop {} {\bibfield  {journal} {\bibinfo  {journal} {Nature Communications}\ }\textbf {\bibinfo {volume} {13}},\ \bibinfo {pages} {5977} (\bibinfo {year} {2022})}\BibitemShut {NoStop}%
\bibitem [{\citenamefont {Xia}\ \emph {et~al.}(2018)\citenamefont {Xia}, \citenamefont {Bian},\ and\ \citenamefont {Kais}}]{xia_electronic_2018}%
  \BibitemOpen
  \bibfield  {author} {\bibinfo {author} {\bibfnamefont {R.}~\bibnamefont {Xia}}, \bibinfo {author} {\bibfnamefont {T.}~\bibnamefont {Bian}},\ and\ \bibinfo {author} {\bibfnamefont {S.}~\bibnamefont {Kais}},\ }\bibfield  {title} {\bibinfo {title} {Electronic structure calculations and the ising hamiltonian},\ }\href@noop {} {\bibfield  {journal} {\bibinfo  {journal} {The Journal of Physical Chemistry B}\ }\textbf {\bibinfo {volume} {122}},\ \bibinfo {pages} {3384} (\bibinfo {year} {2018})}\BibitemShut {NoStop}%
\bibitem [{\citenamefont {Zhang}\ \emph {et~al.}(2017)\citenamefont {Zhang}, \citenamefont {Pagano}, \citenamefont {Hess}, \citenamefont {Kyprianidis}, \citenamefont {Becker}, \citenamefont {Kaplan}, \citenamefont {Gorshkov}, \citenamefont {Gong},\ and\ \citenamefont {Monroe}}]{zhang_dqpt53_2017}%
  \BibitemOpen
  \bibfield  {author} {\bibinfo {author} {\bibfnamefont {J.}~\bibnamefont {Zhang}}, \bibinfo {author} {\bibfnamefont {G.}~\bibnamefont {Pagano}}, \bibinfo {author} {\bibfnamefont {P.~W.}\ \bibnamefont {Hess}}, \bibinfo {author} {\bibfnamefont {A.}~\bibnamefont {Kyprianidis}}, \bibinfo {author} {\bibfnamefont {P.}~\bibnamefont {Becker}}, \bibinfo {author} {\bibfnamefont {H.}~\bibnamefont {Kaplan}}, \bibinfo {author} {\bibfnamefont {A.~V.}\ \bibnamefont {Gorshkov}}, \bibinfo {author} {\bibfnamefont {Z.-X.}\ \bibnamefont {Gong}},\ and\ \bibinfo {author} {\bibfnamefont {C.}~\bibnamefont {Monroe}},\ }\bibfield  {title} {\bibinfo {title} {Observation of a many-body dynamical phase transition with a 53-qubit quantum simulator},\ }\href@noop {} {\bibfield  {journal} {\bibinfo  {journal} {Nature}\ }\textbf {\bibinfo {volume} {551}},\ \bibinfo {pages} {601} (\bibinfo {year} {2017})}\BibitemShut {NoStop}%
\bibitem [{\citenamefont {Wiebe}\ \emph {et~al.}(2016)\citenamefont {Wiebe}, \citenamefont {Kapoor},\ and\ \citenamefont {Svore}}]{wiebe_qdeeplearn_2016}%
  \BibitemOpen
  \bibfield  {author} {\bibinfo {author} {\bibfnamefont {N.}~\bibnamefont {Wiebe}}, \bibinfo {author} {\bibfnamefont {A.}~\bibnamefont {Kapoor}},\ and\ \bibinfo {author} {\bibfnamefont {K.~M.}\ \bibnamefont {Svore}},\ }\bibfield  {title} {\bibinfo {title} {Quantum deep learning},\ }\href@noop {} {\bibfield  {journal} {\bibinfo  {journal} {Quantum Info. Comput.}\ }\textbf {\bibinfo {volume} {16}},\ \bibinfo {pages} {541–587} (\bibinfo {year} {2016})}\BibitemShut {NoStop}%
\bibitem [{\citenamefont {Lin}\ \emph {et~al.}(1993)\citenamefont {Lin}, \citenamefont {Gubernatis}, \citenamefont {Gould},\ and\ \citenamefont {Tobochnik}}]{lin_exact_1993}%
  \BibitemOpen
  \bibfield  {author} {\bibinfo {author} {\bibfnamefont {H.}~\bibnamefont {Lin}}, \bibinfo {author} {\bibfnamefont {J.}~\bibnamefont {Gubernatis}}, \bibinfo {author} {\bibfnamefont {H.}~\bibnamefont {Gould}},\ and\ \bibinfo {author} {\bibfnamefont {J.}~\bibnamefont {Tobochnik}},\ }\bibfield  {title} {\bibinfo {title} {Exact diagonalization methods for quantum systems},\ }\href@noop {} {\bibfield  {journal} {\bibinfo  {journal} {Computers in Physics}\ }\textbf {\bibinfo {volume} {7}},\ \bibinfo {pages} {400} (\bibinfo {year} {1993})}\BibitemShut {NoStop}%
\bibitem [{\citenamefont {White}(1992)}]{white_dmrg_1992}%
  \BibitemOpen
  \bibfield  {author} {\bibinfo {author} {\bibfnamefont {S.~R.}\ \bibnamefont {White}},\ }\bibfield  {title} {\bibinfo {title} {Density matrix formulation for quantum renormalization groups},\ }\href {https://doi.org/10.1103/PhysRevLett.69.2863} {\bibfield  {journal} {\bibinfo  {journal} {Phys. Rev. Lett.}\ }\textbf {\bibinfo {volume} {69}},\ \bibinfo {pages} {2863} (\bibinfo {year} {1992})}\BibitemShut {NoStop}%
\bibitem [{\citenamefont {Vidal}(2004)}]{vidal_efficient_2004}%
  \BibitemOpen
  \bibfield  {author} {\bibinfo {author} {\bibfnamefont {G.}~\bibnamefont {Vidal}},\ }\bibfield  {title} {\bibinfo {title} {Efficient simulation of one-dimensional quantum many-body systems},\ }\href {https://doi.org/10.1103/PhysRevLett.93.040502} {\bibfield  {journal} {\bibinfo  {journal} {Phys. Rev. Lett.}\ }\textbf {\bibinfo {volume} {93}},\ \bibinfo {pages} {040502} (\bibinfo {year} {2004})}\BibitemShut {NoStop}%
\bibitem [{\citenamefont {Haegeman}\ \emph {et~al.}(2011)\citenamefont {Haegeman}, \citenamefont {Cirac}, \citenamefont {Osborne}, \citenamefont {Pi\ifmmode~\check{z}\else \v{z}\fi{}orn}, \citenamefont {Verschelde},\ and\ \citenamefont {Verstraete}}]{haegeman_tdvp_2011}%
  \BibitemOpen
  \bibfield  {author} {\bibinfo {author} {\bibfnamefont {J.}~\bibnamefont {Haegeman}}, \bibinfo {author} {\bibfnamefont {J.~I.}\ \bibnamefont {Cirac}}, \bibinfo {author} {\bibfnamefont {T.~J.}\ \bibnamefont {Osborne}}, \bibinfo {author} {\bibfnamefont {I.}~\bibnamefont {Pi\ifmmode~\check{z}\else \v{z}\fi{}orn}}, \bibinfo {author} {\bibfnamefont {H.}~\bibnamefont {Verschelde}},\ and\ \bibinfo {author} {\bibfnamefont {F.}~\bibnamefont {Verstraete}},\ }\bibfield  {title} {\bibinfo {title} {Time-dependent variational principle for quantum lattices},\ }\href {https://doi.org/10.1103/PhysRevLett.107.070601} {\bibfield  {journal} {\bibinfo  {journal} {Phys. Rev. Lett.}\ }\textbf {\bibinfo {volume} {107}},\ \bibinfo {pages} {070601} (\bibinfo {year} {2011})}\BibitemShut {NoStop}%
\bibitem [{\citenamefont {Schollw\"ock}(2011)}]{schollwoeck_dmrg_2011}%
  \BibitemOpen
  \bibfield  {author} {\bibinfo {author} {\bibfnamefont {U.}~\bibnamefont {Schollw\"ock}},\ }\bibfield  {title} {\bibinfo {title} {The density-matrix renormalization group in the age of matrix product states},\ }\href@noop {} {\bibfield  {journal} {\bibinfo  {journal} {Ann. Physics}\ }\textbf {\bibinfo {volume} {326}},\ \bibinfo {pages} {96 } (\bibinfo {year} {2011})},\ \bibinfo {note} {january 2011 Special Issue}\BibitemShut {NoStop}%
\bibitem [{\citenamefont {Trotter}(1959)}]{trotter_product_1959}%
  \BibitemOpen
  \bibfield  {author} {\bibinfo {author} {\bibfnamefont {H.~F.}\ \bibnamefont {Trotter}},\ }\bibfield  {title} {\bibinfo {title} {On the product of semi-groups of operators},\ }\href@noop {} {\bibfield  {journal} {\bibinfo  {journal} {Proceedings of the American Mathematical Society}\ }\textbf {\bibinfo {volume} {10}},\ \bibinfo {pages} {545} (\bibinfo {year} {1959})}\BibitemShut {NoStop}%
\bibitem [{\citenamefont {Suzuki}(1976)}]{suzuki_generalized_1976}%
  \BibitemOpen
  \bibfield  {author} {\bibinfo {author} {\bibfnamefont {M.}~\bibnamefont {Suzuki}},\ }\bibfield  {title} {\bibinfo {title} {Generalized trotter's formula and systematic approximants of exponential operators and inner derivations with applications to many-body problems},\ }\href@noop {} {\bibfield  {journal} {\bibinfo  {journal} {Communications in Mathematical Physics}\ }\textbf {\bibinfo {volume} {51}},\ \bibinfo {pages} {183} (\bibinfo {year} {1976})}\BibitemShut {NoStop}%
\bibitem [{\citenamefont {Lloyd}(1996)}]{lloyd_universalqsim_1996}%
  \BibitemOpen
  \bibfield  {author} {\bibinfo {author} {\bibfnamefont {S.}~\bibnamefont {Lloyd}},\ }\bibfield  {title} {\bibinfo {title} {Universal quantum simulators},\ }\href@noop {} {\bibfield  {journal} {\bibinfo  {journal} {Science}\ }\textbf {\bibinfo {volume} {273}},\ \bibinfo {pages} {1073} (\bibinfo {year} {1996})}\BibitemShut {NoStop}%
\bibitem [{\citenamefont {Buessen}\ \emph {et~al.}(2023)\citenamefont {Buessen}, \citenamefont {Segal},\ and\ \citenamefont {Khait}}]{buessen_distributed_2023}%
  \BibitemOpen
  \bibfield  {author} {\bibinfo {author} {\bibfnamefont {F.~L.}\ \bibnamefont {Buessen}}, \bibinfo {author} {\bibfnamefont {D.}~\bibnamefont {Segal}},\ and\ \bibinfo {author} {\bibfnamefont {I.}~\bibnamefont {Khait}},\ }\bibfield  {title} {\bibinfo {title} {Simulating time evolution on distributed quantum computers},\ }\href {https://doi.org/10.1103/PhysRevResearch.5.L022003} {\bibfield  {journal} {\bibinfo  {journal} {Phys. Rev. Res.}\ }\textbf {\bibinfo {volume} {5}},\ \bibinfo {pages} {L022003} (\bibinfo {year} {2023})}\BibitemShut {NoStop}%
\bibitem [{\citenamefont {von Keyserlingk}\ \emph {et~al.}(2018)\citenamefont {von Keyserlingk}, \citenamefont {Rakovszky}, \citenamefont {Pollmann},\ and\ \citenamefont {Sondhi}}]{vonkeyserlingk_operator_2018}%
  \BibitemOpen
  \bibfield  {author} {\bibinfo {author} {\bibfnamefont {C.~W.}\ \bibnamefont {von Keyserlingk}}, \bibinfo {author} {\bibfnamefont {T.}~\bibnamefont {Rakovszky}}, \bibinfo {author} {\bibfnamefont {F.}~\bibnamefont {Pollmann}},\ and\ \bibinfo {author} {\bibfnamefont {S.~L.}\ \bibnamefont {Sondhi}},\ }\bibfield  {title} {\bibinfo {title} {Operator hydrodynamics, otocs, and entanglement growth in systems without conservation laws},\ }\href {https://doi.org/10.1103/PhysRevX.8.021013} {\bibfield  {journal} {\bibinfo  {journal} {Phys. Rev. X}\ }\textbf {\bibinfo {volume} {8}},\ \bibinfo {pages} {021013} (\bibinfo {year} {2018})}\BibitemShut {NoStop}%
\bibitem [{\citenamefont {Xu}\ and\ \citenamefont {Swingle}(2024)}]{xu_scrambling_2024}%
  \BibitemOpen
  \bibfield  {author} {\bibinfo {author} {\bibfnamefont {S.}~\bibnamefont {Xu}}\ and\ \bibinfo {author} {\bibfnamefont {B.}~\bibnamefont {Swingle}},\ }\bibfield  {title} {\bibinfo {title} {Scrambling dynamics and out-of-time-ordered correlators in quantum many-body systems},\ }\href {https://doi.org/10.1103/PRXQuantum.5.010201} {\bibfield  {journal} {\bibinfo  {journal} {PRX Quantum}\ }\textbf {\bibinfo {volume} {5}},\ \bibinfo {pages} {010201} (\bibinfo {year} {2024})}\BibitemShut {NoStop}%
\bibitem [{\citenamefont {Schaub}\ \emph {et~al.}(2016)\citenamefont {Schaub}, \citenamefont {O'Clery}, \citenamefont {Billeh}, \citenamefont {Delvenne}, \citenamefont {Lambiotte},\ and\ \citenamefont {Barahona}}]{Schaub2016GraphPartitions}%
  \BibitemOpen
  \bibfield  {author} {\bibinfo {author} {\bibfnamefont {M.~T.}\ \bibnamefont {Schaub}}, \bibinfo {author} {\bibfnamefont {N.}~\bibnamefont {O'Clery}}, \bibinfo {author} {\bibfnamefont {Y.~N.}\ \bibnamefont {Billeh}}, \bibinfo {author} {\bibfnamefont {J.-C.}\ \bibnamefont {Delvenne}}, \bibinfo {author} {\bibfnamefont {R.}~\bibnamefont {Lambiotte}},\ and\ \bibinfo {author} {\bibfnamefont {M.}~\bibnamefont {Barahona}},\ }\bibfield  {title} {\bibinfo {title} {Graph partitions and cluster synchronization in networks of oscillators},\ }\href {https://doi.org/10.1063/1.4961065} {\bibfield  {journal} {\bibinfo  {journal} {Chaos: An Interdisciplinary Journal of Nonlinear Science}\ }\textbf {\bibinfo {volume} {26}},\ \bibinfo {pages} {094821} (\bibinfo {year} {2016})}\BibitemShut {NoStop}%
\bibitem [{\citenamefont {Vidal}\ \emph {et~al.}(1998)\citenamefont {Vidal}, \citenamefont {Mosseri},\ and\ \citenamefont {Dou\ifmmode~\mbox{\c{c}}\else \c{c}\fi{}ot}}]{Vidal1998AharonovBohmCages}%
  \BibitemOpen
  \bibfield  {author} {\bibinfo {author} {\bibfnamefont {J.}~\bibnamefont {Vidal}}, \bibinfo {author} {\bibfnamefont {R.}~\bibnamefont {Mosseri}},\ and\ \bibinfo {author} {\bibfnamefont {B.}~\bibnamefont {Dou\ifmmode~\mbox{\c{c}}\else \c{c}\fi{}ot}},\ }\bibfield  {title} {\bibinfo {title} {Aharonov-bohm cages in two-dimensional structures},\ }\href {https://doi.org/10.1103/PhysRevLett.81.5888} {\bibfield  {journal} {\bibinfo  {journal} {Phys. Rev. Lett.}\ }\textbf {\bibinfo {volume} {81}},\ \bibinfo {pages} {5888} (\bibinfo {year} {1998})}\BibitemShut {NoStop}%
\bibitem [{\citenamefont {Cirino}\ and\ \citenamefont {de~Oliveira}(2026)}]{cirino_connectivity_2026}%
  \BibitemOpen
  \bibfield  {author} {\bibinfo {author} {\bibfnamefont {M.~J.}\ \bibnamefont {Cirino}}\ and\ \bibinfo {author} {\bibfnamefont {M.~C.}\ \bibnamefont {de~Oliveira}},\ }\href {https://arxiv.org/abs/2608.19130} {\bibinfo {title} {Connectivity--interference competition in coherent transport on percolated hierarchical small-world networks}} (\bibinfo {year} {2026}),\ \Eprint {https://arxiv.org/abs/2608.19130} {arXiv:2608.19130 [quant-ph]} \BibitemShut {NoStop}%
\bibitem [{\citenamefont {Rodrigues}\ \emph {et~al.}(2016)\citenamefont {Rodrigues}, \citenamefont {Peron}, \citenamefont {Ji},\ and\ \citenamefont {Kurths}}]{rodrigues_kuramoto_2016}%
  \BibitemOpen
  \bibfield  {author} {\bibinfo {author} {\bibfnamefont {F.~A.}\ \bibnamefont {Rodrigues}}, \bibinfo {author} {\bibfnamefont {T.~K.~D.}\ \bibnamefont {Peron}}, \bibinfo {author} {\bibfnamefont {P.}~\bibnamefont {Ji}},\ and\ \bibinfo {author} {\bibfnamefont {J.}~\bibnamefont {Kurths}},\ }\bibfield  {title} {\bibinfo {title} {The kuramoto model in complex networks},\ }\href {https://doi.org/10.1016/j.physrep.2015.10.008} {\bibfield  {journal} {\bibinfo  {journal} {Physics Reports}\ }\textbf {\bibinfo {volume} {610}},\ \bibinfo {pages} {1} (\bibinfo {year} {2016})}\BibitemShut {NoStop}%
\bibitem [{\citenamefont {Lebrun}\ \emph {et~al.}(2017)\citenamefont {Lebrun}, \citenamefont {Tsunegi}, \citenamefont {Bortolotti}, \citenamefont {Kubota}, \citenamefont {Jenkins}, \citenamefont {Romera}, \citenamefont {Yakushiji}, \citenamefont {Fukushima}, \citenamefont {Grollier}, \citenamefont {Yuasa} \emph {et~al.}}]{lebrun_mutual_2017}%
  \BibitemOpen
  \bibfield  {author} {\bibinfo {author} {\bibfnamefont {R.}~\bibnamefont {Lebrun}}, \bibinfo {author} {\bibfnamefont {S.}~\bibnamefont {Tsunegi}}, \bibinfo {author} {\bibfnamefont {P.}~\bibnamefont {Bortolotti}}, \bibinfo {author} {\bibfnamefont {H.}~\bibnamefont {Kubota}}, \bibinfo {author} {\bibfnamefont {A.}~\bibnamefont {Jenkins}}, \bibinfo {author} {\bibfnamefont {M.}~\bibnamefont {Romera}}, \bibinfo {author} {\bibfnamefont {K.}~\bibnamefont {Yakushiji}}, \bibinfo {author} {\bibfnamefont {A.}~\bibnamefont {Fukushima}}, \bibinfo {author} {\bibfnamefont {J.}~\bibnamefont {Grollier}}, \bibinfo {author} {\bibfnamefont {S.}~\bibnamefont {Yuasa}}, \emph {et~al.},\ }\bibfield  {title} {\bibinfo {title} {Mutual synchronization of spin torque nano-oscillators through a long-range and tunable electrical coupling scheme},\ }\href {https://doi.org/10.1038/ncomms15825} {\bibfield  {journal} {\bibinfo  {journal} {Nature communications}\ }\textbf {\bibinfo {volume} {8}},\ \bibinfo {pages} {15825} (\bibinfo {year}
  {2017})}\BibitemShut {NoStop}%
\bibitem [{\citenamefont {Tsunegi}\ \emph {et~al.}(2018)\citenamefont {Tsunegi}, \citenamefont {Taniguchi}, \citenamefont {Lebrun}, \citenamefont {Yakushiji}, \citenamefont {Cros}, \citenamefont {Grollier}, \citenamefont {Fukushima}, \citenamefont {Yuasa},\ and\ \citenamefont {Kubota}}]{tsunegi_scaling_2018}%
  \BibitemOpen
  \bibfield  {author} {\bibinfo {author} {\bibfnamefont {S.}~\bibnamefont {Tsunegi}}, \bibinfo {author} {\bibfnamefont {T.}~\bibnamefont {Taniguchi}}, \bibinfo {author} {\bibfnamefont {R.}~\bibnamefont {Lebrun}}, \bibinfo {author} {\bibfnamefont {K.}~\bibnamefont {Yakushiji}}, \bibinfo {author} {\bibfnamefont {V.}~\bibnamefont {Cros}}, \bibinfo {author} {\bibfnamefont {J.}~\bibnamefont {Grollier}}, \bibinfo {author} {\bibfnamefont {A.}~\bibnamefont {Fukushima}}, \bibinfo {author} {\bibfnamefont {S.}~\bibnamefont {Yuasa}},\ and\ \bibinfo {author} {\bibfnamefont {H.}~\bibnamefont {Kubota}},\ }\bibfield  {title} {\bibinfo {title} {Scaling up electrically synchronized spin torque oscillator networks},\ }\href {https://doi.org/10.1038/s41598-018-31769-9} {\bibfield  {journal} {\bibinfo  {journal} {Scientific reports}\ }\textbf {\bibinfo {volume} {8}},\ \bibinfo {pages} {13475} (\bibinfo {year} {2018})}\BibitemShut {NoStop}%
\bibitem [{\citenamefont {Kozachkov}\ \emph {et~al.}(2025)\citenamefont {Kozachkov}, \citenamefont {Slotine},\ and\ \citenamefont {Krotov}}]{kozachkov_neuronastrocyte_2025}%
  \BibitemOpen
  \bibfield  {author} {\bibinfo {author} {\bibfnamefont {L.}~\bibnamefont {Kozachkov}}, \bibinfo {author} {\bibfnamefont {J.-J.}\ \bibnamefont {Slotine}},\ and\ \bibinfo {author} {\bibfnamefont {D.}~\bibnamefont {Krotov}},\ }\bibfield  {title} {\bibinfo {title} {Neuron–astrocyte associative memory},\ }\href {https://doi.org/10.1073/pnas.2417788122} {\bibfield  {journal} {\bibinfo  {journal} {Proceedings of the National Academy of Sciences}\ }\textbf {\bibinfo {volume} {122}},\ \bibinfo {pages} {e2417788122} (\bibinfo {year} {2025})}\BibitemShut {NoStop}%
\bibitem [{\citenamefont {Kaiser}\ and\ \citenamefont {Datta}(2021)}]{kaiser_probabilistic_2021}%
  \BibitemOpen
  \bibfield  {author} {\bibinfo {author} {\bibfnamefont {J.}~\bibnamefont {Kaiser}}\ and\ \bibinfo {author} {\bibfnamefont {S.}~\bibnamefont {Datta}},\ }\bibfield  {title} {\bibinfo {title} {Probabilistic computing with p-bits},\ }\href {https://doi.org/10.1063/5.0067927} {\bibfield  {journal} {\bibinfo  {journal} {Applied Physics Letters}\ }\textbf {\bibinfo {volume} {119}},\ \bibinfo {pages} {150503} (\bibinfo {year} {2021})}\BibitemShut {NoStop}%
\end{thebibliography}%

\end{document}